\documentclass[useAMS,usenatbib,amsmath,amssymb,referee]{mn2e}
\usepackage{graphicx}
\usepackage{bm}
\usepackage{longtable}
\usepackage{dcolumn}
\usepackage{amsmath}
\usepackage{amssymb}
\usepackage{hyperref}
\usepackage{mathrsfs}                %Ó¢ÎÄ»¨Ìå
\usepackage{multirow}

\title[]{Cosmological test using strong gravitational lensing systems}

\author[]{C. C. Yuan$^{1}$, F. Y. Wang$^{1,2}$
\thanks{fayinwang@nju.edu.cn(FYW)}\\
$^{1}$School of Astronomy and Space Science, Nanjing University,
Nanjing
210093, China\\
$^{2}$Key Laboratory of Modern Astronomy and Astrophysics (Nanjing
University), Ministry of Education, Nanjing 210093, China }

\begin{document}
\maketitle
\begin{abstract}
As one of the probes of universe, strong gravitational lensing
systems allow us to compare different cosmological models and
constrain vital cosmological parameters. This purpose can be reached
from the dynamic and  geometry properties of strong gravitational
lensing systems, for instance, time-delay $\Delta\tau$ of images,
the velocity dispersion $\sigma$ of the lensing galaxies and the
combination of these two effects, $\Delta\tau/\sigma^2$. In this
paper, in order to carry out one-on-one comparisons between
$\Lambda$CDM universe and $R_h=ct$ universe, we use a sample
containing 36 strong lensing systems with the measurement of
velocity dispersion from the SLACS and LSD survey. Concerning the
time-delay effect, 12 two-image lensing systems with $\Delta\tau$
are also used. In addition, Monte Carlo (MC) simulations are used to
compare the efficiency of the three methods as mentioned above. From
simulations, we estimate the number of lenses required to rule out
one model at the $99.7\%$ confidence level. Comparing with
constraints from $\Delta\tau$ and the velocity dispersion $\sigma$,
we find that using $\Delta\tau/\sigma^2$ can improve the
discrimination between cosmological models. Despite the
independence tests of these methods reveal a correlation between
$\Delta\tau/\sigma^2$ and $\sigma$, $\Delta\tau/\sigma^2$ could be
considered as an improved method of $\sigma$ if more data samples
are available.
\end{abstract}
\begin{keywords}
gravitational lensing - Galaxy - cosmology: dark energy
\end{keywords}

%%%%%%%%%%%%%%%%%%%%%%%%%%%%%% section 1, introduction %%%%%%%%%%%%%%%%%%%%%%%%%%%%%
\section{Introduction}
Detailed study of type Ia supernovae
\citep{Riess1998,Perlmutter1999} has revealed that our universe is
undergoing an era of accelerating expansion, which suggests the
composition of our universe may include some unknown components such
as dark energy. Observations from other independent methods such as
cosmic microwave background (CMB), baryon acoustic oscillations
(BAO), clusters of galaxies, gamma-ray bursts (Wang et al. 2015) and
large-scale structure can lead to the same result. The cosmological
constant $(\Lambda)$ is considered to be the best candidate of dark
energy, which is accordant with many observations
\citep{Riess2004,Davis2007,Kowalski2008,Wang2011,Wang2014}. However,
there are many other models that were proposed to explain the
observations, one promising model is $R_h=ct$ model
\citep{Melia2007,Melia2012,Yu2014}. The existence of so many
theoretical models calls for more precise and complementary data to
differentiate between the models.

Since Walsh et al. (1979) discovered the strong gravitational
lensing in Q0957+561, strong gravitational lensing has become one
powerful probe in the study of cosmology
\citep{Zhu2000a,Chae2003b,Chae2004,Mitchell2004,Zhu2008a,Zhu2008b}
and astrophysics, i.e., measuring the mass of galaxies or clusters.
Up to now, hundreds of lensing systems produced by galaxies
or quasars have been discovered, but only one part of them with
geometry and dynamic information can be used for statistical
analysis. The observations of about 70 lensing systems provide the
data required not only for studying the statistical properties of
galaxy structures and mass distribution \citep{Ofek2003,Chae2003},
but also for confining cosmological parameters. The Einstein radius
obtained from the deflection angle and the time-delay of different
images can provide the information of angular diameter
distance (i.e. $D_{ds}$ and $D_s$) independently, and further can
be used to constrain cosmological models.

In recent years, many tests based on strong gravitational lensing
have been used to constrain cosmological parameters. For example,
the statistical data in Cosmic Lens All-Sky Survey (CLASS)
demonstrated $\Omega_{\rm m}\approx0.3$ assuming a flat cosmology
and non-evolving galaxy populations \citep{Chae2003b}. Assuming a
mean galaxy density profile that does not evolve with redshift, a
$\Lambda$-dominated cold dark matter cosmology, and Gaussian
distributions for bulk parameters describing the lens and source
populations, Dobke et al. (2009) found a sample of $\sim$400
time-delay lenses can reach the similar levels of precision as from
the best of other methods. Coe \& Moustakas (2009) presented the
first analysis of time-delay lenses to constrain a broad range of
cosmological parameters. Using the 80 $D_{ds}/D_s$ data from various
gravitational lens survey and lensing galaxy cluster with X-ray
observations and optical giant luminous arcs, Cao et al. (2012)
obtained $\Omega_{\rm m}=0.20^{+0.07}_{-0.07}$ in the $\Lambda$CDM
model. The application of selection methods in one-on-one
comparisons between $\Lambda$CDM and $R_h=ct$ Universe has found
that the former is favored by the data \citep{Melia2014}. In their
simulations of velocity dispersion $\sigma$, in order to rule out
$R_h=ct$ at the $99.7\%$ confidence level assuming the cosmology is
$\Lambda$CDM, they found about 200 lens systems are required, while
a sample of at least 300 systems to rule out $\Lambda$CDM if the
background is $R_h=ct$. Similar results were obtained in the
simulations of $\Delta\tau$ \citep{Wei2014}. However, Paraficz \&
Hjorth (2009) argues that $\Delta\tau/\sigma^2$ is more effective to
constrain cosmological parameters than $\Delta\tau$ and $\sigma$
separately.

In this paper, we focus on constraining cosmological parameters
using observational data of $\sigma$ and $\Delta\tau$ in a
sample containing 36 lensing and a sample of 12
time-delays. In addition, we perform one-on-one comparisons between
the $R_h=ct$ model and the $\Lambda$CDM model through MC simulations
of $\Delta\tau/\sigma^2$, $\Delta\tau$ and $\sigma^2$ to estimate
the number of data points needed to rule out one model in the
background of another at the $99.7\%$ confidence level. To
achieve this goal we assume the three methods are independent and
the dependence tests are performed later.

The paper is organised as follows. In the next section, we
introduce the strong gravitational lensing systems as the probe of
the universe. In section 3, we test the $\Lambda$CDM and $R_h=ct$
models utilising the measured data samples. In section 4,
we use MC simulations of $\Delta\tau$, $\sigma$ and the combination
$\Delta\tau/\sigma^2$ independently to perform one-on-one
comparisons. We also compare the capability of these three methods
and test their independence. Conclusions and discussions
are given in section 5.

%%%%%%%%%%%%%%%%%%%%%%%%%%%%%%%%%%%%%%%%%%%%%%%%%% section 2 %%%%%%%%%%%%%%%%%%%%%%%%%%%%%%%%%%%%%%%
\section{Strong Lenses as a Probe of the Universe}
In this paper, we mainly concern two cosmological models: the
$\Lambda$CDM and the $R_h=ct$ models. In the $\Lambda$CDM model,
angular diameter depends on several parameters, including Hubble
constant $H_0$, density fractions
$\Omega_{\rm{m}}=\rho_{\rm{m}}/\rho_{\rm{c}}$,
$\Omega_{\rm{r}}=\rho_{\rm{r}}/\rho_{\rm{c}}$ and
$\Omega_{\Lambda}=\rho_{\Lambda}/\rho_{\rm{c}}$, where
$\rho_{\rm{m}}$, $\rho_{\rm{r}}$ and $\rho_{\Lambda}$ are current
matter, radiation, dark energy densities respectively, and
$\rho_{\rm{c}}=3c^2H_0^2/8\pi G$ is critical density of our
universe. Assuming a zero spatial curvature universe, we have
$\Omega_{\rm{m}}+\Omega_{\rm{r}}+\Omega_{\Lambda}=1$. The angular
diameter distance between redshifts $z_1$ and $z_2 (>z_1)$ is given
by the formula
\begin{equation}
D^{\Lambda\rm{CDM}}(z_1,z_2)=\frac{c}{H_0}\frac{1}{1+z_2}\int_{z_l}^{z_2}\left[\Omega_{\rm{m}}(1+z)^3+\Omega_{\rm{r}}(1+z)^4+\Omega_{\Lambda}\right].
\end{equation}
Since radiation is insignificant at gravitational lensing redshifts
and noting $\Omega_{\rm{m}}+\Omega_{\Lambda}=1$, we have two
essential parameters needed to be constrained, including $H_0$ and
$\Omega_{\rm{m}}$. In the $R_h=ct$ model
\citep{Melia2007,Melia2012}, there is only one parameter $H_0$ in
the angular diameter distance
\begin{equation}
D^{\rm{R_h=ct}}(z_1,z_2)=\frac{c}{H_0}\frac{1}{1+z_2}{\rm{ln}}\left(\frac{1+z_2}{1+z_1}\right).
\end{equation}

Strong gravitational lensing occurs when the observer, the lens and
the source are well aligned that we can get separate images of the
source due to the gravitational field of the lens. The time-delay
$\Delta\tau$ is caused by the difference in length of the optical
paths and the gravitational time dilation for the ray passing
through the effective gravitational potential of the lens
$\Psi(\vec\theta_i)$. For a given image $i$ at angle position
$\vec\theta_i$ with the source position at angle $\vec\beta$, time
delay $\Delta\tau_i$ can be written as (Blandford \& Narayan 1986)
\begin{equation}
\Delta\tau_i=\frac{1+z_l}{c}\frac{D_{OS}D_{OL}}{D_{LS}}\left[\frac{1}{2}(\vec{\theta_i}-\vec{\beta})^2-\Psi(\vec{\theta_i})\right],
\end{equation}
where $z_l$ is the redshift of the lens, $D_{OL}, D_{OS}, D_{LS}$ are
the angular diameter distances between observer and lens, observer
and source, and lens and source, respectively. If the lens geometry
$\vec{\theta_i}-\vec{\beta}$ and the effective gravitational
potential of the lens $\Psi(\vec\theta_i)$ are known, we can define
the time-delay distance
\begin{equation}
\mathcal{D}^{\rm{time-delay}}(z_l,z_s)=\frac{D_{OS}D_{OL}}{D_{LS}}.
\end{equation}
If such systems have only two images at $\vec\theta_A$ and
$\vec\theta_B$, the time delay is given by the expression
\begin{equation}
\Delta\tau=\frac{1+z_l}{2c}\mathcal{D}^{\rm{time-delay}}(z_l,z_s)({\vec\theta_B}^2-{\vec\theta_A}^2),
\end{equation}
under the single isothermal sphere (SIS) model.

Another method to constrain cosmological models is to use the Einstein radius in the SIS model,
\begin{equation}
\theta_E=4\pi\frac{D_{LS}}{D_{OS}}\frac{\sigma_{SIS}^2}{c^2},
\end{equation}
which varies with cosmological models via the ratio of
angular diameter distances between lens/source, and
observer/source. From equations (3) and (4), we can see that
time-delay is proportional to $D_{OL}D_{OS}/D_{LS}$ and the square
of the velocity dispersion is proportional to
$D_{OS}/D_{LS}$. The ratio $\Delta\tau/\sigma^2$ is determined only
by $D_{OL}$, that is to say,
\begin{equation}
\Delta\tau\propto\frac{D_{OS}D_{OL}}{D_{LS}},\sigma_{SIS}^2\propto\frac{D_{OS}}{D_{LS}},\frac{\Delta\tau}{\sigma_{SIS}^2}\propto {D_{OL}}.
\end{equation}
We show the relations between the redshift of lens $(z_l)$ and the
three quantities in the equation (7) in $\Lambda$CDM model with a
fixed source redshift $z_s=3$. In Figure 1, we plot these quantities
in several cases relative to the Einstein-de Sitter Universe
($\Omega_{\rm m}=1, \Omega_{\Lambda}=0$) as in Paraficz \& Hjorth
(2009). The extent of separations between curves in Figure 1 reveals
the sensitivity of the corresponding method to discriminate
cosmological models. Comparing with constraints from $\Delta\tau$
and the velocity dispersion $\sigma$, we find that using
$\Delta\tau/\sigma^2$ can significantly improve the discrimination
between cosmological models. Meanwhile, the sensitivity increases
with the redshift of lens, thus, it is of special significance to
study high-redshift lenses.

For simplicity, we still follow the approximation in Paraficz \&
Hjorth (2009) that $\theta_E=(\theta_A+\theta_B)/2$ and
$\theta_B>\theta_A$. From equations (5) and (6), we obtain
\begin{equation}
D_{OL}(\theta_B-\theta_A)=\frac{c^3}{4\pi}\frac{\Delta\tau}{\sigma_{SIS}^2(1+z_l)}.
\end{equation}

Up to now, we have three methods: velocity dispersion $\sigma$, time
delay $\Delta\tau$ and the combination $\Delta\tau/\sigma^2$. The
relations between these quantities and angular distances can be
found in equations (5), (6) and (8).  Equations about strong
gravitational lensing in our paper are based on the SIS model.
However, Treu et al. (2006) found that the ratio between the
velocity dispersion $\sigma_0$ of the lensing galaxy and the
velocity dispersion $\sigma_{SIS}$ for the corresponding singular
isothermal sphere or ellipsoid, $\sigma_0/\sigma_{SIS}$, is close to
unity. Here, we assume $\sigma_{SIS}=f_{E}\sigma_0$.
\begin{figure}
\begin{center}
\includegraphics[width=0.9\textwidth]{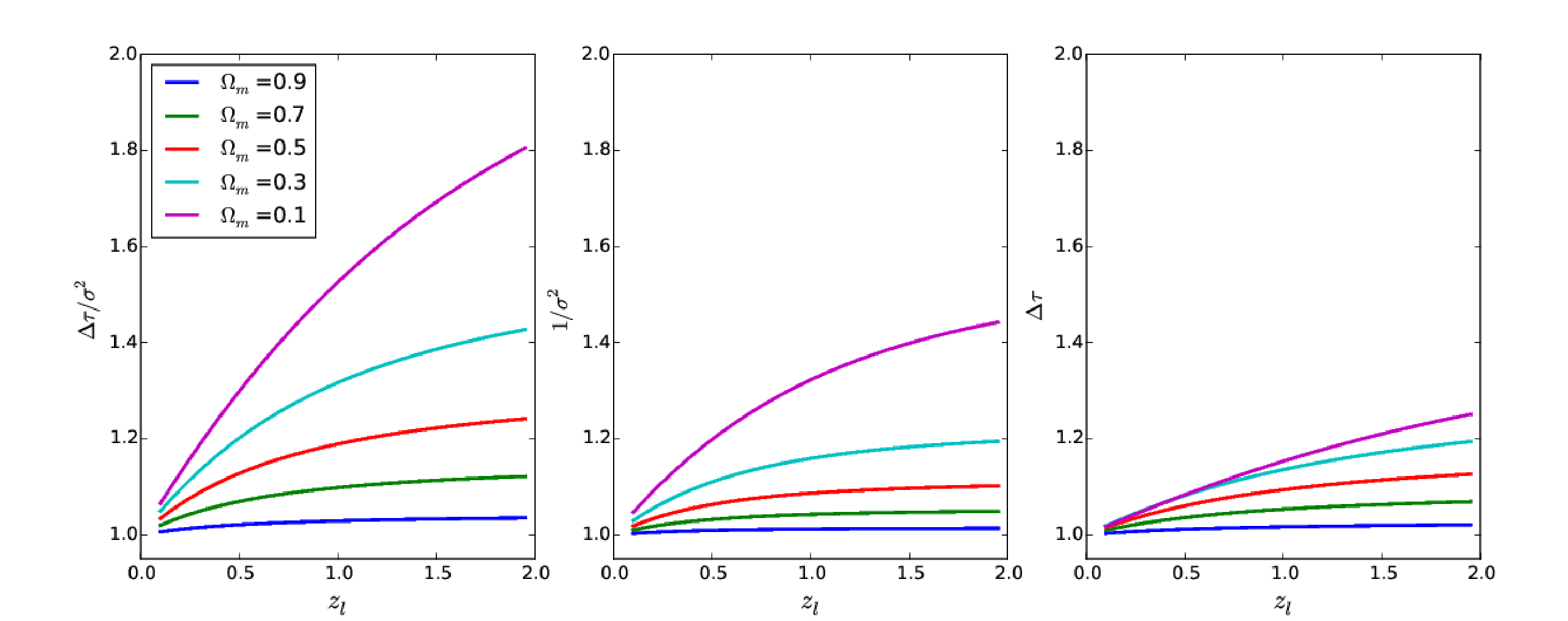}
\caption{Sensitivity of three methods($\sigma$, $\Delta\tau$ and
$\Delta\tau/\sigma^2$) to compare models\textbf{(also see Paraficz
\& Hjorth 2009)}. The source redshift $z_l$ is fixed to 3. We
compare five $\Omega_{\rm m}$ values: 0.1, 0.3, 0.5,0.7,0.9. Each is
obtained relative to the Einstein-de Sitter Universe.}
\end{center}
\end{figure}

%%%%%%%%%%%%%%%%%%%%%%%%%%%%%%%%%%%%%%%%%%%%%%%%%%%%%%%% section 3, Samples and results %%%%%%%%%%%%%%%%%%%%%%%%%%%%%%%%%%%%%%%%%%%%%%%%%%%
\section{Samples and results}

\subsection{Samples Used}
In consideration of the consistency with a simple power-law (or even
SIS) profile, SIS lens should have only 2 images
\citep{Biesiada2010,Biesiada2011}, thus we use 36 lenses which have
only two images in sample I for Einstein ring data. All lenses in
our sample had well measured central dispersions taken from the
SLACS and LSD surveys \citep{Biesiada2010,Bolton2008,Newton2011}. We
use time-delay lenses to compare cosmological models as well. In our
paper, 12 time-delay lensing systems are contained in sample II.

For each model, we find the best fit by minimizing the $\chi^2$
function
\begin{equation}
\chi^2=\sum_{i}\frac{(\mathcal{D}_{i}^{th}-\mathcal{D}_i^{obs})^2}{\sigma^2_{\mathcal{D},i}},
\end{equation}
where $\mathcal{D}^{th}=D_{LS}/D_{OS}$ and
$\mathcal{D}^{th}=D_{OS}D_{OL}/D_{LS}$ for Einstein circle lenses
(ECL) and time-delay lenses (TDL), respectively. In $\chi^2$
function, $\sigma_{\mathcal{D},i}$ donates the variance of
$\mathcal{D}_i^{obs}$ and it can be obtained from the error propagation
formula of $\mathcal{D}^{obs}$. Thus, the standard deviation of
$\mathcal{D}_i^{obs}$ from Einstein circle lensing is
\begin{equation}
\sigma_{ECL}=\mathcal{D}^{obs}\left[\left(\frac{\sigma_{\theta_E}}{\theta_E}\right)^2+4\left(\frac{\sigma_{\sigma_0}}{\sigma_0}\right)^2+4\left(\frac{\sigma_{f_E}}{f_E}\right)^2\right],
\end{equation}
and we take $5\%$ error both for $f_E$ (Grillo et al. 2008) and
$\theta_E$. The standard deviation of time-delay can be written as

\begin{equation}
\sigma_{TDL}=\mathcal{D}^{obs}\left[\left(\frac{\sigma_{\Delta\tau}}{\Delta\tau}\right)^2+4\left(\frac{\theta_B\sigma_{\theta_B}}{\theta_B^2-\theta_A^2}\right)^2
+4\left(\frac{\theta_A\sigma_{\theta_A}}{\theta_B^2-\theta_A^2}\right)^2+\eta^2\right].
\end{equation}
Here, we also introduce the parameter $\eta$, to represent the
derivation from the SIS model.

%%%%%%%%%%%%%%%%%%%%%%%%%%%%%%%%%%%%%%%%%%%%%%%% subsection %%%%%%%%%%%%%%%%%%%%%%%%%%%%%%%%%%%%%%%%%%%%%
\subsection{Cosmological Models Test}
At first, we use 36 lenses in sample I (summarised in Table 1) to
compare $\Lambda$CDM with $R_h=ct$ Universe. In this case, there are
two free parameters ($\Omega_{\rm m}$ and $f_E$) in $\Lambda$CDM and
one parameter ($f_E$) in $R_h=ct$ model. Using sample I, we find the
minimum $\chi^2$=45.6 for $f_E=1.007^{+0.023}_{-0.028}(1\sigma)$,
$\Omega_{\rm{m}}=0.15^{+0.243}_{-0.144}(1\sigma)$ in the
$\Lambda$CDM. For the $R_h=ct$ Universe, the best fit is
$f_E=1.03^{+0.035}_{-0.028}$ at 1$\sigma$ confidence level. The
results are shown in Figure 2. In order to compare the constraints,
the relation between $\mathcal{D}^{obs}$ and $\mathcal{D}^{th}$ for
the best-fitting parameters is shown in Figure 3. Using 36 lensing
systems we find the $\chi^2$ values for both $\Lambda$CDM and
$R_h=ct$ are high. Obviously, the parameters are not well
constrained. The primary reason is that the number of data points is
too small to yield a good constraint.

%%%%%%%%%%%%%%%%%%%%%%%%%%%%%%%%%%%%%%%%%%%%%%%%%%%% Table 1 and Table 2

\begin{table}
\centering
\begin{tabular}{cccccc}
\hline
\hline
System & $z_l$ &$z_s$ & $\theta_E$(arcsec) & $\sigma_0({\rm km/s})$ & Refs\\
\hline
SDSS J0037-0942 &   0.1955  &   0.6322  &   1.47    &   $   282 \pm 11  $    &  1       \\
SDSS J0216-0813 &   0.3317  &   0.5235  &   1.15    &   $   349 \pm 24  $    &  1       \\
SDSS J0737+3216 &   0.3223  &   0.5812  &   1.03    &   $   326 \pm 16  $    &  1       \\
SDSS J0912+0029 &   0.1642  &   0.3239  &   1.61    &   $   325 \pm 12  $    &  1       \\
SDSS J1250+0523 &   0.2318  &   0.795   &   1.15    &   $   274 \pm 15  $    &  1       \\
SDSS J1630+4520 &   0.2479  &   0.7933  &   1.81    &   $   279 \pm 17  $    &  1       \\
SDSS J2300+0022 &   0.2285  &   0.4635  &   1.25    &   $   305 \pm 19  $    &  1       \\
SDSS J2303+1422 &   0.1553  &   0.517   &   1.64    &   $   271 \pm 16  $    &  1       \\
CFRS03.1077 &   0.938   &   2.941   &   1.24    &   $   251 \pm 19  $   &    1      \\
HST 15433   &   0.497   &   2.092   &   0.36    &   $   116 \pm 10  $   &    1      \\
MG 2016 &   1.004   &   3.263   &   1.56    &   $   328 \pm 32  $   &   1        \\
SDSS J0955+0101 &   0.1109  &   0.3159  &   0.91    &   $   192 \pm 13  $    &  2       \\
SDSS J0959+4416 &   0.2369  &   0.5315  &   0.96    &   $   244 \pm 19  $    &  2       \\
SDSS J1143-0144 &   0.106   &   0.4019  &   1.68    &   $   269 \pm 13  $    &  2       \\
SDSS J1205+4910 &   0.215   &   0.4808  &   1.22    &   $   281 \pm 14  $    &  2       \\
SDSS J1403+0006 &   0.1888  &   0.473   &   0.83    &   $   213 \pm 17  $    &  2       \\
SDSS J1403+0006 &   0.1888  &   0.473   &   0.83    &   $   213 \pm 17  $    &  2     \\
SDSS J0044+0113 &   0.1196  &   0.1965  &   0.79    &   $   266 \pm 13  $    &  2,3     \\
SDSS J0330-0020 &   0.3507  &   1.0709  &   1.1 &   $   212 \pm 21  $   &    2,3        \\
SDSS J0935-0003 &   0.3475  &   0.467   &   0.87    &   $   396 \pm 35  $    &  2,3     \\
SDSS J1112+0826 &   0.273   &   0.6295  &   1.49    &   $   320 \pm 20  $    &  2,3     \\
SDSS J1142+1001 &   0.2218  &   0.5039  &   0.98    &   $   221 \pm 22  $    &  2,3     \\
SDSS J1204+0358 &   0.1644  &   0.6307  &   1.31    &   $   267 \pm 17  $    &  2,3     \\
SDSS J1213+6708 &   0.1229  &   0.6402  &   1.42    &   $   292 \pm 15  $    &  2,3     \\
SDSS J1218+0830 &   0.135   &   0.7172  &   1.45    &   $   219 \pm 11  $    &  2,3     \\
SDSS J1432+6317 &   0.123   &   0.6643  &   1.26    &   $   199 \pm 10  $    &  2,3     \\
SDSS J1436-0000 &   0.2852  &   0.8049  &   1.12    &   $   224 \pm 17  $    &  2,3     \\
SDSS J1443+0304 &   0.1338  &   0.4187  &   0.81    &   $   209 \pm 11  $    &  2,3     \\
SDSS J1451-0239 &   0.1254  &   0.5203  &   1.04    &   $   223 \pm 14  $    &  2,3     \\
SDSS J1525+3327 &   0.3583  &   0.7173  &   1.31    &   $   264 \pm 26  $    &  2,3     \\
SDSS J1531-0105 &   0.1596  &   0.7439  &   1.71    &   $   279 \pm 14  $    &  2,3     \\
SDSS J1538+5817 &   0.1428  &   0.5312  &   1   &   $   189 \pm 12  $   &    2,3        \\
SDSS J1621+3931 &   0.2449  &   0.6021  &   1.29    &   $   236 \pm 20  $    &  2,3     \\
SDSS J2238-0754 &   0.1371  &   0.7126  &   1.27    &   $   198 \pm 11  $    &  2,3     \\
Q0957+561   &   0.36    &   1.41    &   1.41    &   $   167 \pm 10  $   &    4      \\
MG1549+3047 &   0.11    &   1.17    &   1.15    &   $   227 \pm 18  $   &    5      \\
CY2201-3201 &   0.32    &   3.9 &   0.41    &   $   130 \pm 20  $   &    6,7,8      \\

\hline

\end{tabular}
\caption{The 36 two-image lensing systems in sample I. References:
1. Biesiada, Piorkowska \& Malec (2010); 2. Bolto et al. (2008); 3.
Newton et al. (2011); 4. Young et al. (1980); 5. Leh$\rm \acute{a}$r et al.
(1993); 6. Koopmans \& Treu (2002); 7. Koopmans \& Treu (2003); 8.
Trey \& Koppmans (2004). }
\end{table}

\begin{table}
\centering
\begin{tabular}{ccccccc}
\hline
\hline
System & $z_l$ & $z_s$ & $\theta_A$(arcsec) & $\theta_B$(arcsec) & $\Delta t=t_A-t_B$(days) & Refs\\
\hline
B0218+357 & 0.685 &0.944 &0.057 $\pm$ 0.004& 0.280 $\pm$ 0.008& +10.5 $\pm$ 0.2&  1,2,3\\
B1600+434& 0.414& 1.589 &1.14 $\pm$ 0.075 &0.25 $\pm$ 0.074 &-51.0 $\pm$ 2.0&4,5\\
FBQ0951+2635& 0.26 &1.246& 0.886 $\pm$ 0.004& 0.228 $\pm$ 0.008& -16.0 $\pm$ 2.0& 6\\

HE1104-1805& 0.729 &2.319& 1.099 $\pm$ 0.004 &2.095 $\pm$ 0.008 &+152.2 $\pm$ 3.0 & 2,7,8\\
HE2149-2745& 0.603 &2.033& 1.354 $\pm$ 0.008 &0.344 $\pm$ 0.012& -103.0 $\pm$ 12.0& 6,9\\
PKS1830-211& 0.89 &2.507 &0.67 $\pm$ 0.08 &0.32 $\pm$ 0.08& -26 $\pm$ 5 &10,11\\
Q0142-100& 0.49& 2.719 &1.855 $\pm$ 0.002& 0.383 $\pm$ 0.005 &-89 $\pm$ 11&6,12\\
Q0957+561& 0.36 &1.413& 5.220 $\pm$ 0.006 &1.036 $\pm$ 0.11& -417.09 $\pm$ 0.07&13,14\\
SBS 0909+532& 0.83 &1.377& 0.415 $\pm$ 0.126& 0.756 $\pm$ 0.152& +45.0 $\pm$ 5.5&6,15\\
SBS 1520+530 &0.717 &1.855& 1.207 $\pm$ 0.004& 0.386 $\pm$ 0.008 &-130.0 $\pm$ 3.0 &6,16\\
SDSS J1206+4332& 0.748& 1.789 &1.870 $\pm$ 0.088 &1.278 $\pm$ 0.097& -116 $\pm$ 5 & 17\\
SDSS J1650+4251& 0.577& 1.547 &0.872 $\pm$ 0.027 &0.357 $\pm$ 0.042 &-49.5 $\pm$ 1.9 & 6,18\\

\hline
\end{tabular}
\caption{Time-delay (two-image) lenses in sample II. References: 1.
Carilli et al. (1993); 2. Leh$\rm\acute{a}$r et al. (2000); 3.
Wucknitz et al. (2004); 4. Jackson et al. (1995); 5. Dai \& Kochanek
(2005); 6. Kochanek et al. (2008); 7. Wisotzki et al. (1993); 8.
Poindexter et al. (2007); 9. Burud et al. (2002); 10. Lovell et al.
(1998); 11. Meylan et al. (2005); 12. Koptelova et al. (2012); 13.
Falco et al. (1997); 14. Colley et al. (2003); 15. Dai \& Kochanek
(2009); 16. Auger et al. (2008); 17. Paraficz et al. (2009); 18.
Vuissoz et al. (2007).}
\end{table}

%%%%%%%%%%%%%%%%%%%%%%%%%%%%%%%%%%%%%%%%%%%%%%%%%%%% figure 2, \Lambda CDM constrain %%%%%%%%%%%%%%%%%%%
\begin{figure}
%\begin{center}
{\includegraphics[width=0.5\textwidth]{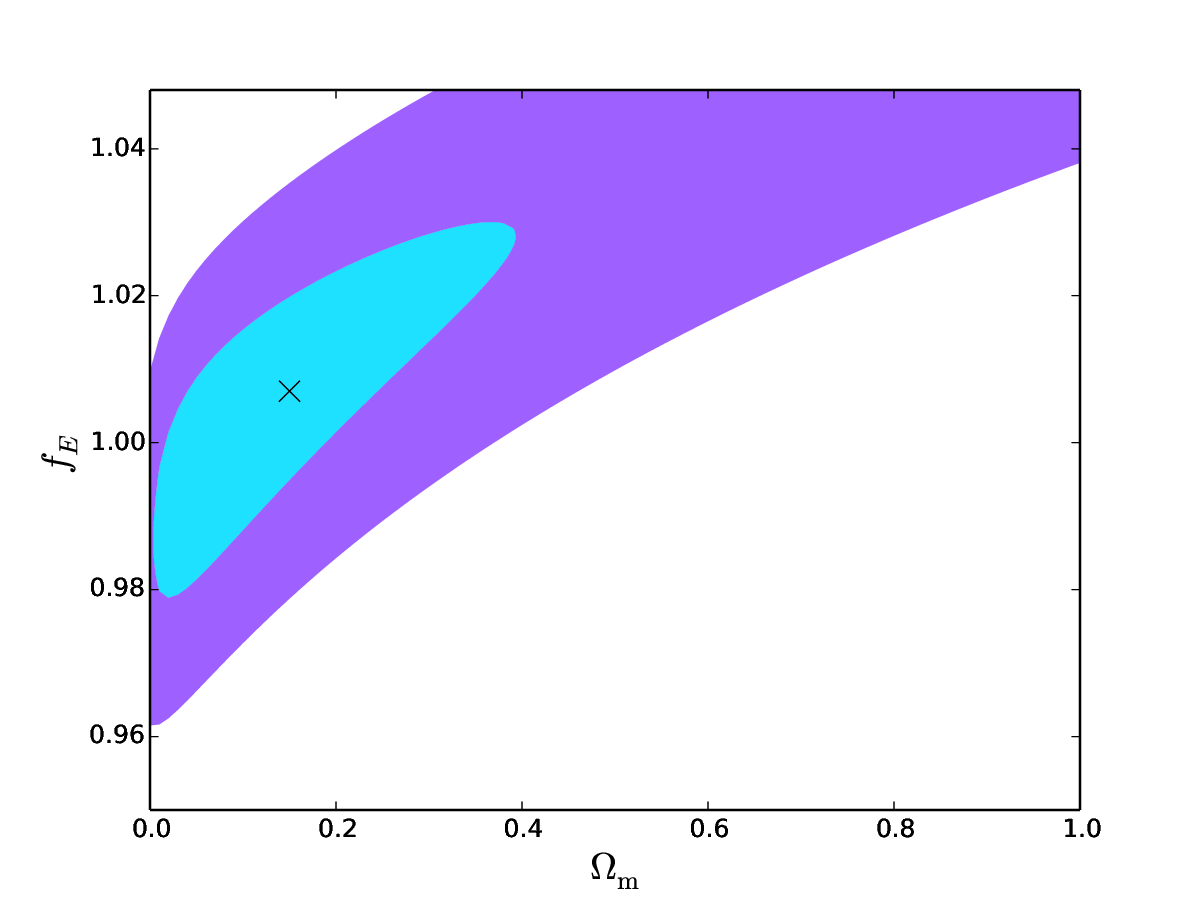}}
{\includegraphics[width=0.5\textwidth]{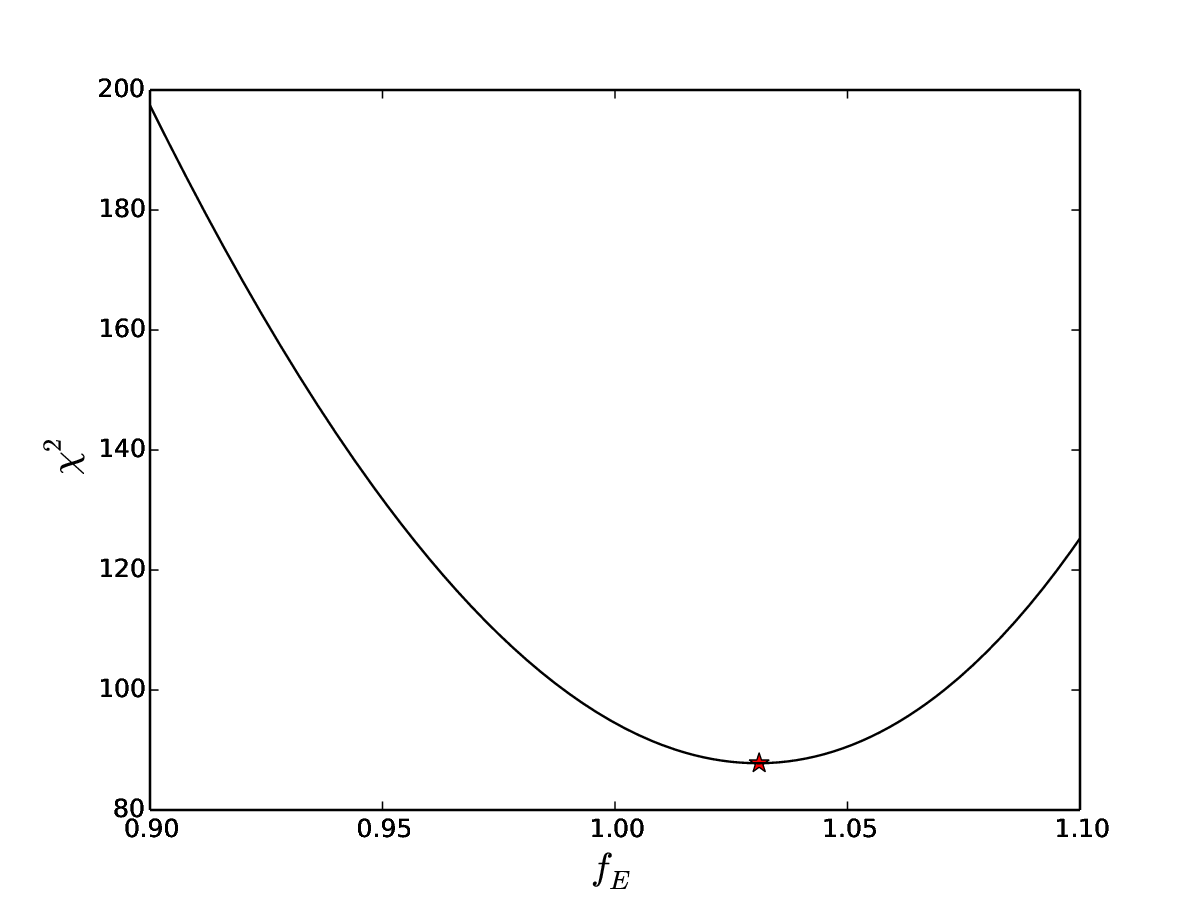}} \caption{The 68.3\%
and 95.4\% confidence region for $\Lambda$CDM model in the
$\Omega_{\rm{m}}-f_E$ plane (left) and the value of $\chi^2$ as a
function of $f_E$ in $R_h=ct$ model (right). The cross in left panel
represents the best-fitting point in $\Lambda$CDM with
$\chi^2_{min}=45.6$ and the star in right panel is the best fit in
$R_h=ct$ with $\chi^2_{min}=48.7$.}
%\end{center}
\end{figure}
%%%%%%%%%%%%%%%%%%%%%%%%%%%%%%%%%%%% figure 3, D^obs vs D^th %%%%%%%%%%%%%%%%%%%%%%%%%%%%%%%%%%%%%%%%%%%
\begin{figure}
%\begin{center}
{\includegraphics[width=0.5\textwidth]{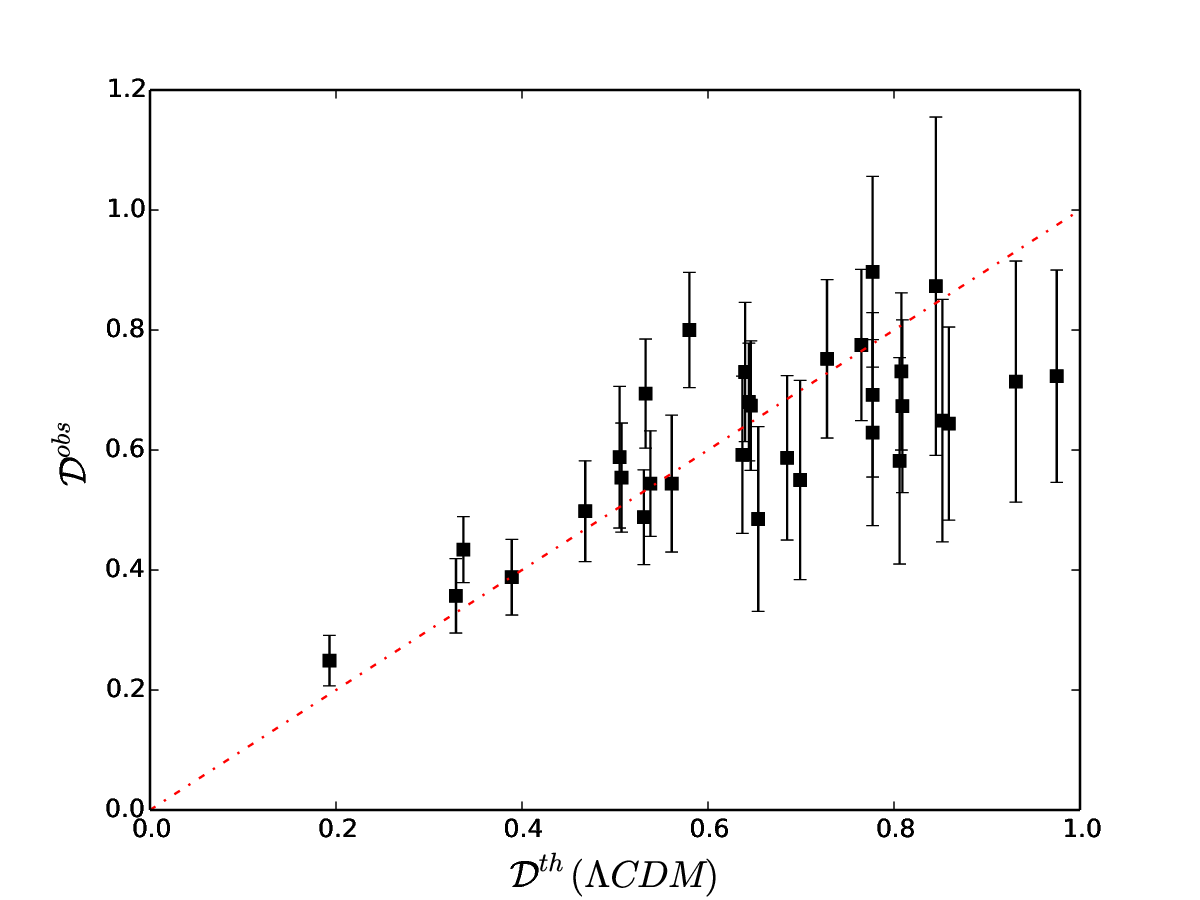}}
{\includegraphics[width=0.5\textwidth]{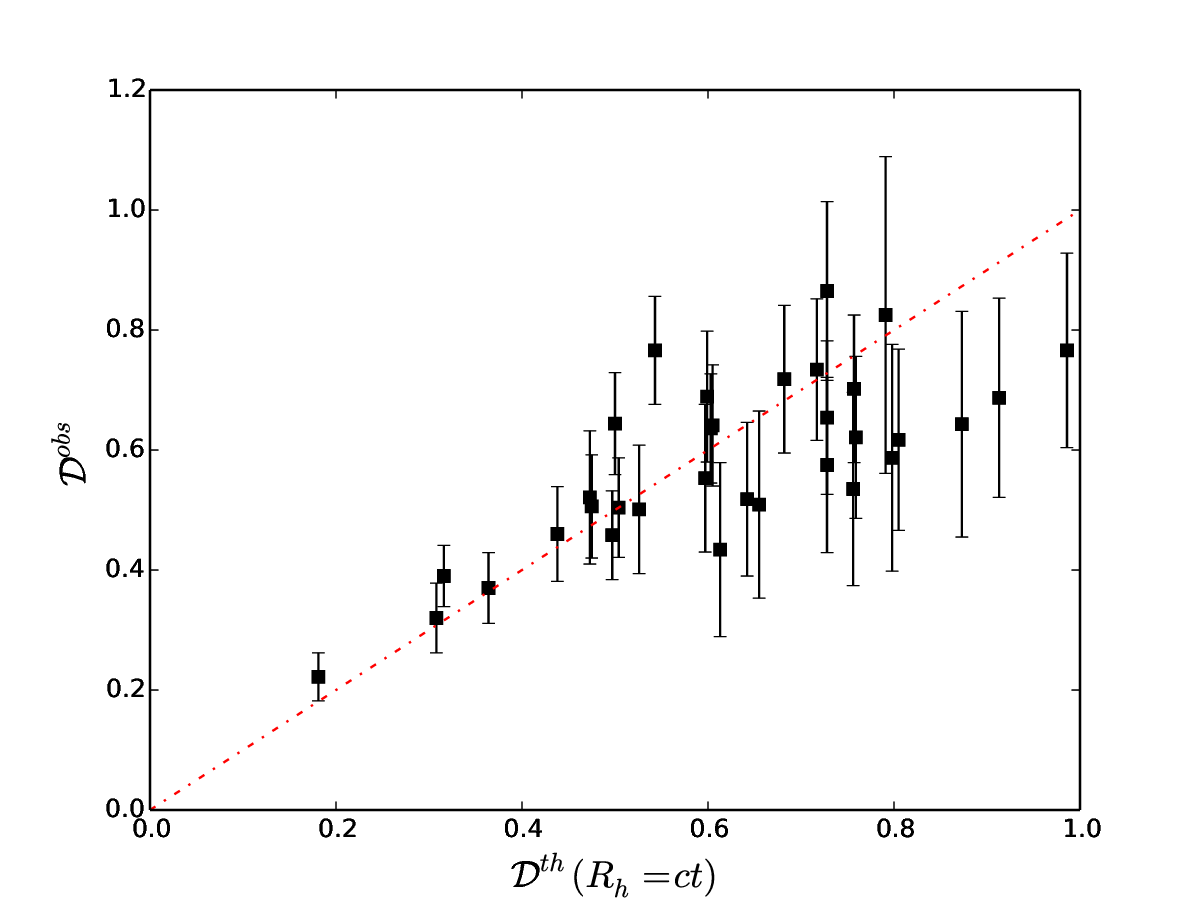}} \caption{Thirty-six
$\mathcal{D}^{obs}$ measurements with error bars, comparisons
between two theoretical models: $\Lambda$CDM (left) and $R_h=ct$
(right).}
%\end{center}
\end{figure}

In addition, we use 12 time-delay lensing systems in sample II
(summarised in Table 2) to compare these two models using
$\mathcal{D}=D_{OS}D_{OL}/D_{LS}$. In the $\Lambda$CDM model, there
are three parameters ($\Omega_{\rm m}, H_0, \eta$) and two
parameters ($H_0, \eta$) in $R_h=ct$ Universe to be constrained.
From maximizing the likelihood function
\begin{equation}
\mathcal{L}\propto\Pi_{i}\frac{1}{\sqrt{\sigma_{TDL,i}}}{\rm
exp}(-\chi^2_i/2),
\end{equation}
we obtain the best-fitting parameters: $\Omega_{\rm
m}=0.19^{+0.24}_{-0.16}(1\sigma), H_0=86.8^{+5}_{-4}~{\rm km\
s^{-1}\ Mpc^{-1}}(1\sigma),\eta=0.28^{+0.03}_{-0.02}(1\sigma)$ in
the $\Lambda$CDM model and $H_0=80.5^{+4}_{-3}{\rm km\ s^{-1}\
Mpc^{-1}},\eta=0.29^{+0.02}_{-0.02}$ in the $R_h=ct$ model. Here, we
marginalize $H_0$ in the $\Lambda$CDM model to find the confidence
levels in $\Omega_{\rm m}-\eta$ plane,
\begin{equation}
\mathcal{L}(\Omega_{\rm m},\eta)=\int{\mathcal{L}(\Omega_{\rm
m},H_0,\eta)}P(H_0){\rm{d}}H_0,
\end{equation}
where $P(H_0)$ is the probability distribution of $H_0$. Figure 4
presents the constraints on $\Omega_{\rm m}-\eta$ plane and
$H_0-\eta$ plane. Similar as in the Einstein circle lensing systems
test, we compare $\mathcal{D}^{obs}$ and $\mathcal{D}^{th}$ in
Figure 5.
%%%%%%%%%%%%%%%%%%%%%%%%%%%%%%%%%%%%%%%%%%%%%% figure 4 %%%%%%%%%%%%%%%%%%%%%%%%%%%%%%555
\begin{figure}
%\begin{center}
{\includegraphics[width=0.5\textwidth]{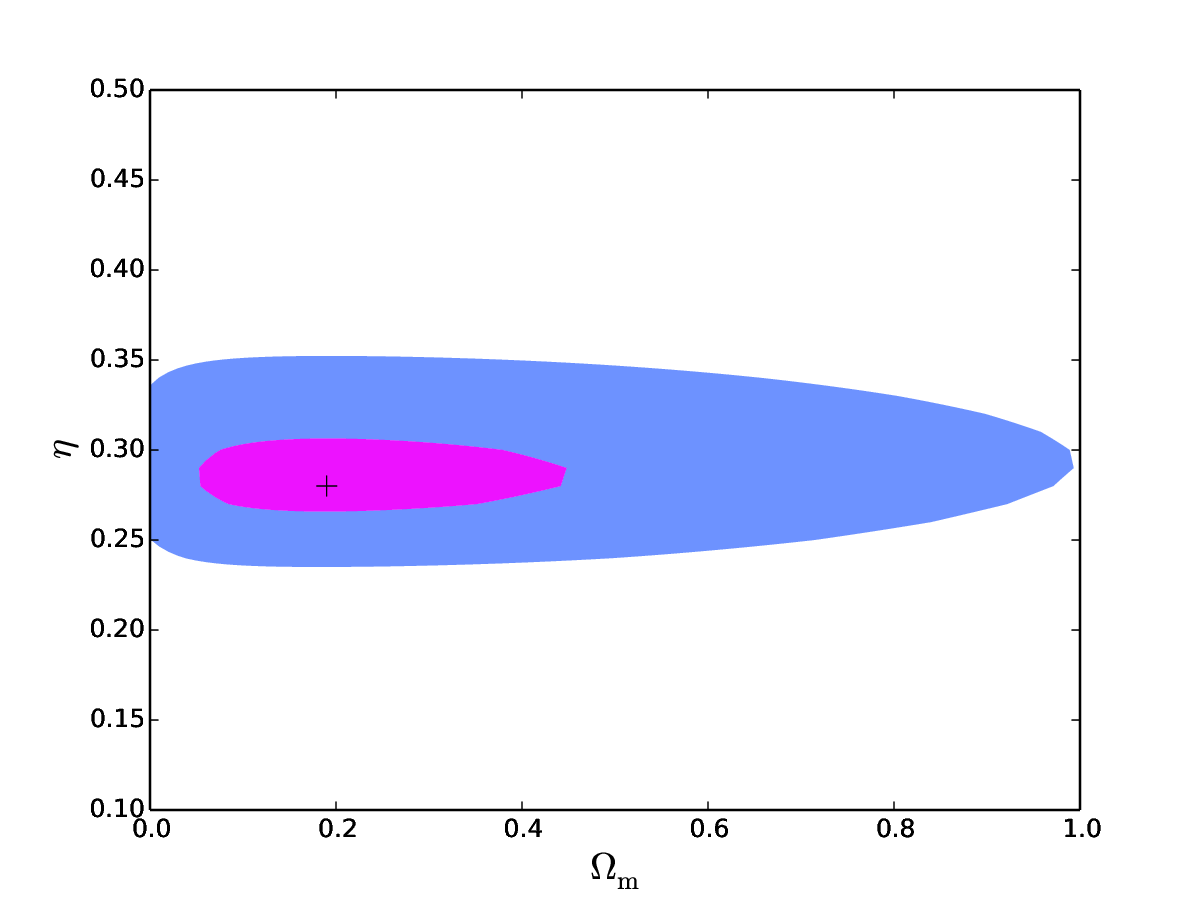}}
{\includegraphics[width=0.5\textwidth]{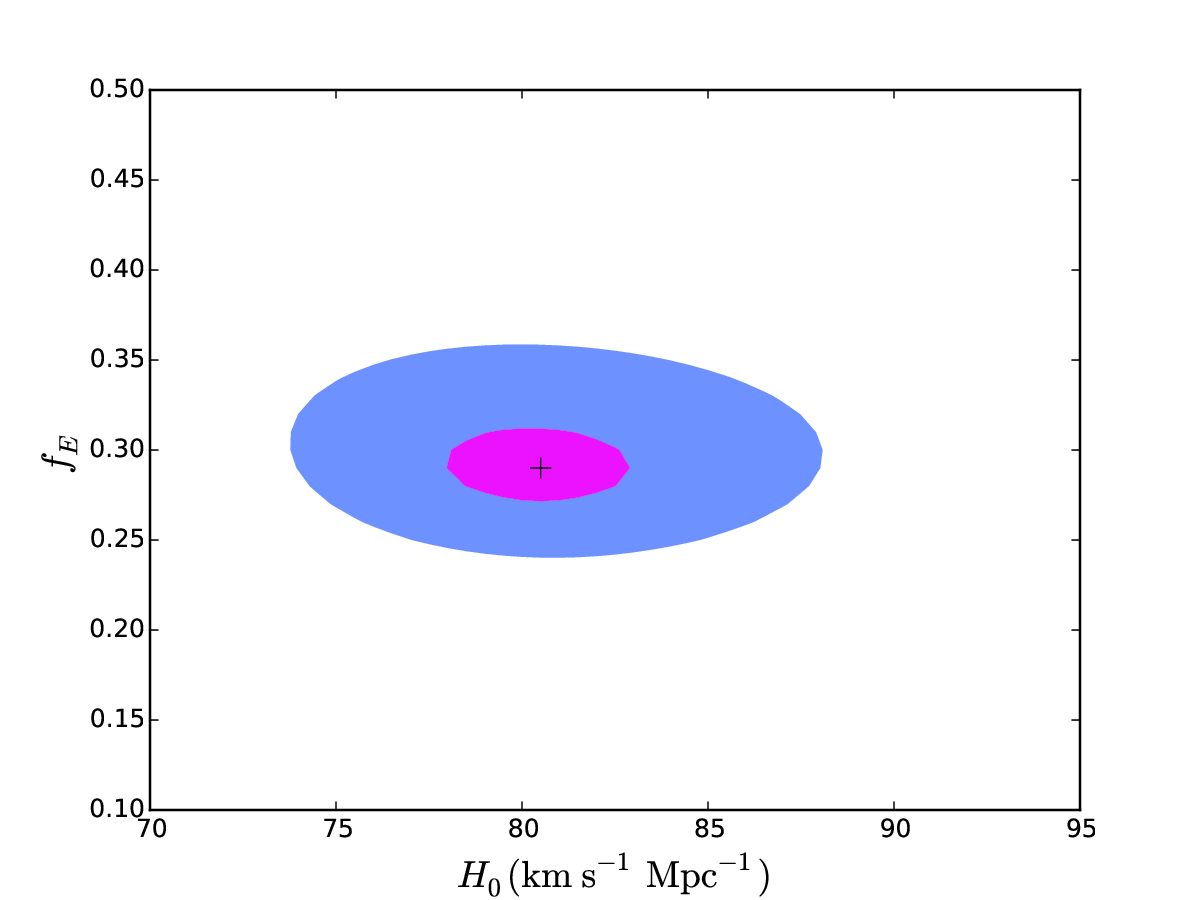}} \caption{The 68.3\%
and 95.4\% confidence region for $\Lambda$CDM model in the
$\Omega_{\rm{m}}-\eta$ plane and the for $R_h=ct$ model in
$H_0-\eta$ plane. The crosses represent the best fit points.}
\end{figure}
%%%%%%%%%%%%%%%%%%%%%%%%%%%%%%%%%%%%%%%%%%%%%%%5 figure 5, D^obs vs D^th  %%%%%%%%%%%%%%%%%%%%%%%%%%%%
\begin{figure}
%\begin{center}
{\includegraphics[width=0.5\textwidth]{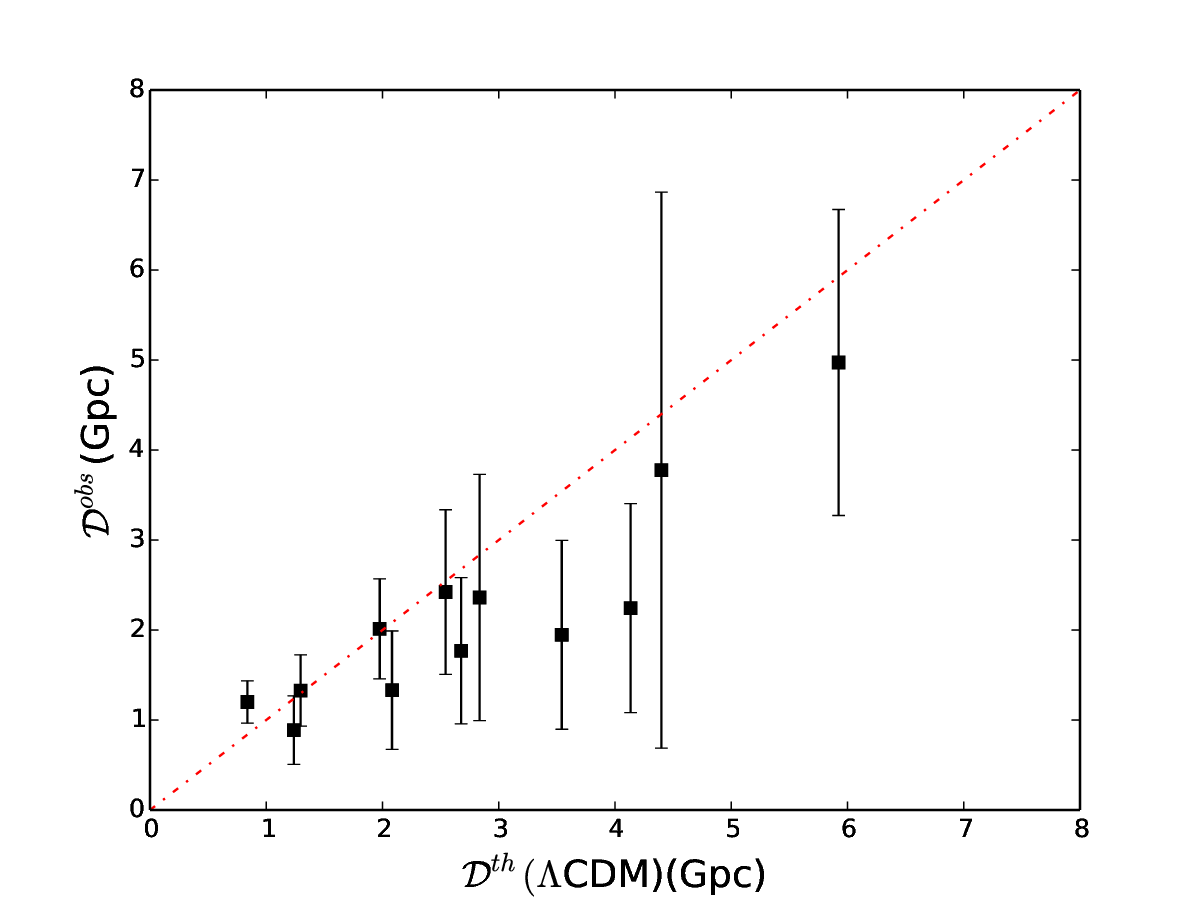}}
{\includegraphics[width=0.5\textwidth]{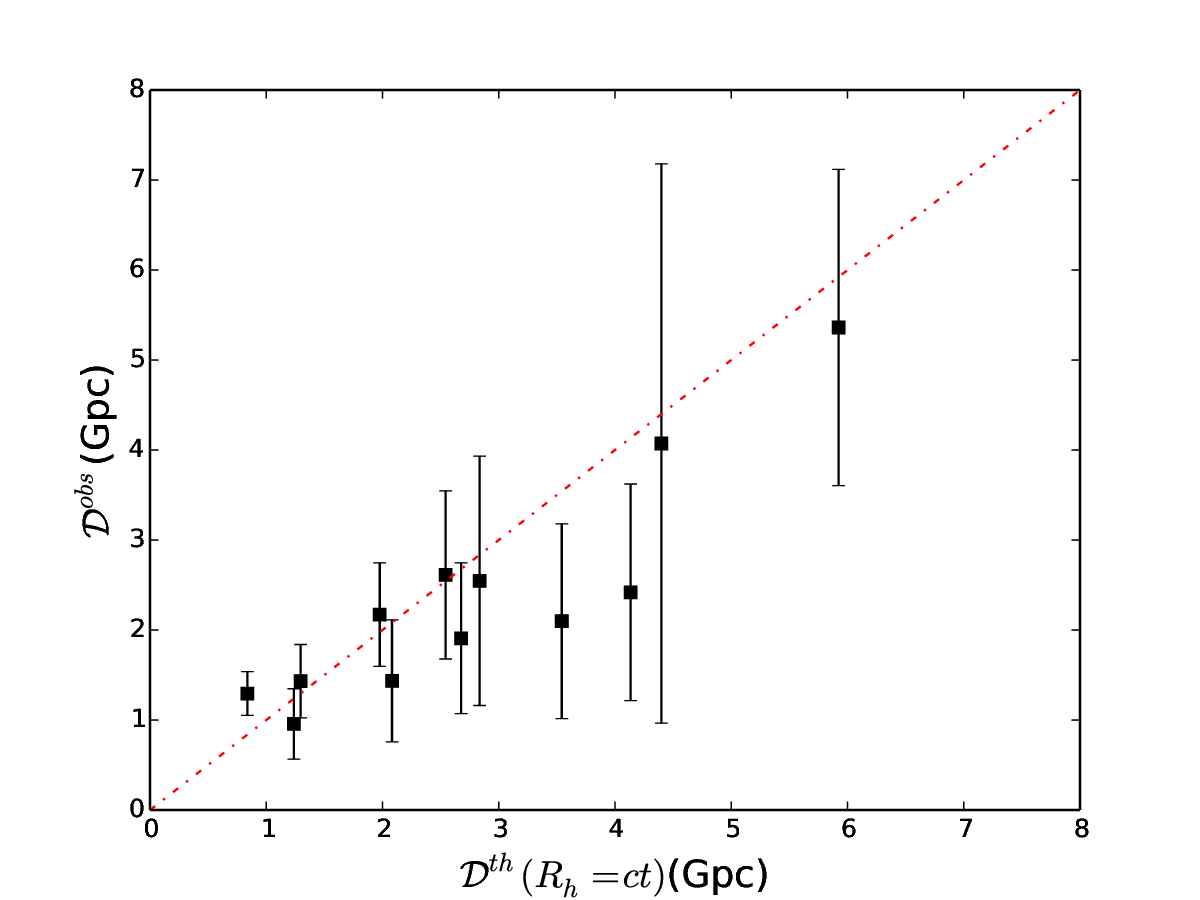}} \caption{Twelve
time-delay $\mathcal{D}^{obs}$ measurements with error bars,
compared with two theoretical models: $\Lambda$CDM (left) and
$R_h=ct$ (right).}
%\end{center}
\end{figure}

To compare two models using Einstein circle lenses, we calculate the
reduced $\chi^2_r$, which is defined as the ratio of the minimum of
$\chi^2$ and the degree of freedom. The degrees of freedom are
$36-2=34$ for $\Lambda$CDM and $36-1=35$ for $R_h=ct$. Thus, we
obtain $(\chi^2_r)_{\Lambda {\rm CDM}}=1.34$ and
$(\chi^2_r)_{R_h=ct}=1.39$. Now, the difference in $\chi^2_r$
between these two models is not big enough to provide strong
evidence which model is better than another. In the time-delay
lensing test, we use the Akaike Information Criterion, ${\rm
AIC}=-2{\rm{ln}}\mathcal{L}+2n$, where $\mathcal{L}$ is the maximum
likelihood and $n$ is the number of free parameters \citep{Liddle2007}.
For the model $\mathcal{M}_\alpha(\alpha=1,2$, i.e. $R_h=ct$ and
$\Lambda$CDM respectively) with ${\rm AIC}_\alpha$, the likelihood
can be written as
\begin{equation}
\mathcal{P}(\mathcal{M}_{\alpha})=\frac{{\rm{exp}}(-{\rm AIC}_{\alpha}/2)}{{\rm{exp}}(-{\rm AIC}_{1}/2)+{\rm{exp}}(-{\rm AIC}_{2}/2)}.
\end{equation}
Then we get ${\rm AIC}_{{\Lambda}{\rm CDM}}=9.35$, ${\rm
AIC}_{R_h=ct}=7.71$, and $\mathcal{P}(\mathcal{M}_1)=0.748$. So the
likelihood of $R_h=ct$ being correct is $74.8\%$ and for
$\Lambda$CDM, the corresponding probability is $25.2\%$. In order to
rule out one model at a $99.7\%$ confidence level, samples
containing more data points are required.

%%%%%%%%%%%%%%%%%%%%%%%%%%%%%%%%%%%%%%%%%%%%%%%%% section 4, MCMC simulations %%%%%%%%%%%%%%%%%%%%%%%%%%%%%%%%%%%%%%%%%%
\section{MC simulations with a Mock Sample}
We make one-on-one comparisons between $\Lambda$CDM and $R_h=ct$
using three different observed quantities ($\sigma$, $\Delta\tau$
and $\Delta\tau/\sigma^2$) under the assumption that these
methods are independent. In our paper, we estimate the number of
lenses needed by using different methods to rule out another model
at the $99.7\%$ confidence level. From the values of $z_l$, $z_s$
and $\sigma$ in observed lensing data, in our simulations, the
redshift of sources are equally distributed between $1.2$ to $3.0$
and the lens redshift between $0.1$ to $1.0$. In the first method
(using $\sigma^2$), we assume the velocity dispersions are uniformly
distributed from $100$ to $300\ \rm{km\ s^{-1}}$
\citep{Paraficz2009}. Then we infer $\theta_E$ from equation (6)
with $f_E=1.0$. We assign the uncertainty of $\theta_E$ to be $5\%$.
Since the simulated sample contains a large number of data points,
the Bayes Information Criterion (${\rm BIC}$) is more appropriate
\begin{equation}
{\rm BIC}=-2{\rm{ln}}\mathcal{L}+({\rm ln}N)n=\chi^2+({\rm ln}N)n,
\end{equation}
where $N$ and $n$ are the number of data points and free parameters
\citep{Schwarz1978}. The form of the likelihood here is similar with
equation (14), where ${\rm AIC}_i$ is substituted by ${\rm BIC}_i$.
Then, we estimate the number of data points needed to rule
out one model (i.e. $R_h=ct$) using another model (i.e.
$\Lambda$CDM) as the background universe.

In the simulations of time-delay lensing systems (using
$\Delta\tau$), we assume the time-delays are uniformly distributed
between -150 to 150 days and we then infer
${\Theta}=\theta_B^2-\theta_A^2$ from equation (5). We assume the
uncertainty of $\Theta$ is $5\%$. In the simulations of the
combination of $\Delta\tau$ and $\sigma^2$, the distributions of
$\Delta\tau$ and $\sigma$ are same with the first two methods and
$\Delta\theta=\theta_B-\theta_A$ is inferred from equation (8). We
still assign the uncertainty of $5\%$ to $\Delta\theta$. The
parameters to be constrained in different models and methods are
summarized in Table 3. We assume $\Omega_{\rm m}=0.3,
H_0=70~\rm{km\ s^{-1}\ Mpc^{-1}}$ and $H_0=70~\rm{km\ s^{-1}\
Mpc^{-1}}$ for the $\Lambda$CDM background and the $R_h=ct$
background, respectively.

%%%%%%%%%%%%%%%%%%%%%%%%%%%%%%%%%%%%%%%% table 3%%%%%%%%%%%%%%%%%%%%%
\begin{table}
\begin{center}
\begin{tabular}{c|c|c|c}
\hline
Method &Model & Free parameters& Degree of freedom\\
\hline
\multirow{2}{1.2cm}{$\sigma$}&{$\Lambda$CDM}&$\Omega_{\rm m}, f_E$&2\\
{} &$R_h=ct$&$f_E$&1\\
\hline
\multirow{2}{1.2cm}{$\Delta\tau$}&{$\Lambda$CDM}&$\Omega_{\rm m},H_0$ & 2\\
{} &$R_h=ct$&$H_0$&1\\
\hline
\multirow{2}{1.2cm}{$\Delta\tau/\sigma^2$}&{$\Lambda$CDM}&$\Omega_{\rm m}, H_0$&2\\
{} &$R_h=ct$&$H_0$&1\\
\hline
\end{tabular}
\caption{The parameters to be constrained in different models and methods.}
\end{center}
\end{table}
%%%%%%%%%%%%%%%%%%%%%%%%%%%%%%%%%%%%% table 2%%%%%%%%%%%%%%%%%%%%%%

Since $\Omega_{\rm{m}}$ is the mutual parameter in three different
methods, in order to differentiate these methods, we study the constraints
on $\Omega_{\rm m}$ using 200 simulated data points in both
$\Lambda$CDM and $R_h=ct$ backgrounds. This simulation is repeated
for 1000 times to find the statistical distributions of the
best-fitting $\Omega_{\rm m}$.

One general concern is whether the $\Delta\tau/\sigma^2$ method is
independent with the other methods since it is derived from $\sigma$
and $\Delta\tau$. We will discuss this issue in the subsection 4.3.

%%%%%%%%%%%%%%%%%%%%%%%%%%%%%%%% subsection %%%%%%%%%%%%%%%%%%%%%%%
\subsection{$\Lambda$CDM Background Cosmology}
In this case, we assume the background universe is $\Lambda$CDM and
seek for the least number of data points needed to rule out $R_h=ct$
at a $99.7\%$ confidence level. We find samples of 300, 200 and 150
data points are needed utilising $\sigma$, $\Delta\tau$ and
$\Delta\tau/\sigma^2$ respectively. The constraints on parameters
and $\rm BIC$ are listed in Table 4. The $68.3\%$ and $95.4\%$
confidence regions for $\Lambda$CDM model in $\Omega_{\rm m}-f_E$
plane for the $\sigma$ method and $\Omega_{\rm m}-H_0$ plane for both
$\Delta\tau$ and $\Delta\tau/\sigma^2$ methods are illustrated in
Figure 6. In figure 7, we show the $\chi^2$ distribution for
parameters ($f_E$ and $H_0$) in the $R_h=ct$ model.

From Table 4 we find that a sample of less data points is needed
using the $\Delta\tau/\sigma^2$ method comparing with the methods of
$\sigma$ and $\Delta\tau$. From Figure 6, we find that the
constraints on different parameters vary with methods. For example,
the method of $\Delta\tau$ is more favorable to constrain the Hubble
constant ($H_0$), while $\Delta\tau/\sigma^2$ can constrain
$\Omega_{\rm m }$ better. The cross between the contour plots of the
$\Delta\tau$ method and $\Delta\tau/\sigma^2$ in Figure 6 shows that
the combination of these two methods can give tighter constraints on
both $H_0$ and $\Omega_{\rm m}$. This conclusion can be
confirmed from the inset of Figure 6 (right) with the best fitting
parameters: $\Omega_{\rm m}=0.30^{+0.05}_{-0.03}$,
$H_0=69.9^{+0.4}_{-0.6}{\rm km\ s^{-1}\ Mpc^{-1}}$. A credible
comparison between three different methods can be obtained from our
1000 repetitive simulations. In this simulation, each sample for
these methods contains 200 simulated data points and we run 1000
minimizations for each method respectively. The distributions of
optimal $\Omega_{\rm m}$ are shown in Figure 8. In order to
differentiate these three methods quantitatively, we use normal
distribution function to fit the distributions of optimal
$\Omega_{\rm m}$ and we find the FWHMs are 0.157, 0.093 and 0.084
for the methods of $\sigma$, $\Delta\tau$, $\Delta\tau/\sigma^2$
respectively. Thus the constraint on $\Omega_{\rm m}$ using
$\Delta\tau/\sigma^2$ is tighter than other two methods.
%%%%%%%%%%%%%%%%%%%%%%%%%%%%%%%%%%% table 4%%%%%%%%%%%%%%%%%%%%%%%%%%%
\begin{table}
\begin{center}

\begin{tabular}{c|c|c|c}
\hline
Method &Model & Best-fitting parameter$(1\sigma)$& $\rm BIC$\\
\hline
\multirow{2}{2.2cm}{$\sigma$($N=300$)}&{$\Lambda$CDM}&$\Omega_{\rm m}=0.36^{+0.18}_{-0.14}, f_E=0.993^{+0.019}_{-0.013}$&339\\
{} &$R_h=ct$&$f_E=0.982^{+0.015}_{0.011}$&357\\
\hline
\multirow{2}{2.2cm}{$\Delta\tau(N=200)$}&{$\Lambda$CDM}&$\Omega_{\rm m}=0.32_{-0.28}^{+0.11},H_0=69.8_{-1.7}^{+0.8} {\rm km\ s^{-1}\ Mpc^{-1}}$ & 238\\
{} &$R_h=ct$&$H_0=69.4_{-0.9}^{+0.7} {\rm km\ s^{-1}\ Mpc^{-1}}$&256\\
\hline
\multirow{2}{2.8cm}{$\Delta\tau/\sigma^2(N=150)$}&{$\Lambda$CDM}&$\Omega_{\rm m}=0.31{\pm0.08}, H_0=69.4\pm3.6 {\rm km\ s^{-1}\ Mpc^{-1}}$&169\\
{} &$R_h=ct$&$H_0=64.1_{-1.2}^{+1.0} {\rm km\ s^{-1}\ Mpc^{-1}}$&187\\
\hline
\end{tabular}
\caption{Results of one-on-one model comparisons in $\Lambda$CDM background.}
\end{center}
\end{table}
%%%%%%%%%%%%%%%%%%%%%%%%%%%%%%%%%% figure 6 %%%%%%%%%%%%%%%%%%%%%5
\begin{figure}
\includegraphics[width=0.5\textwidth]{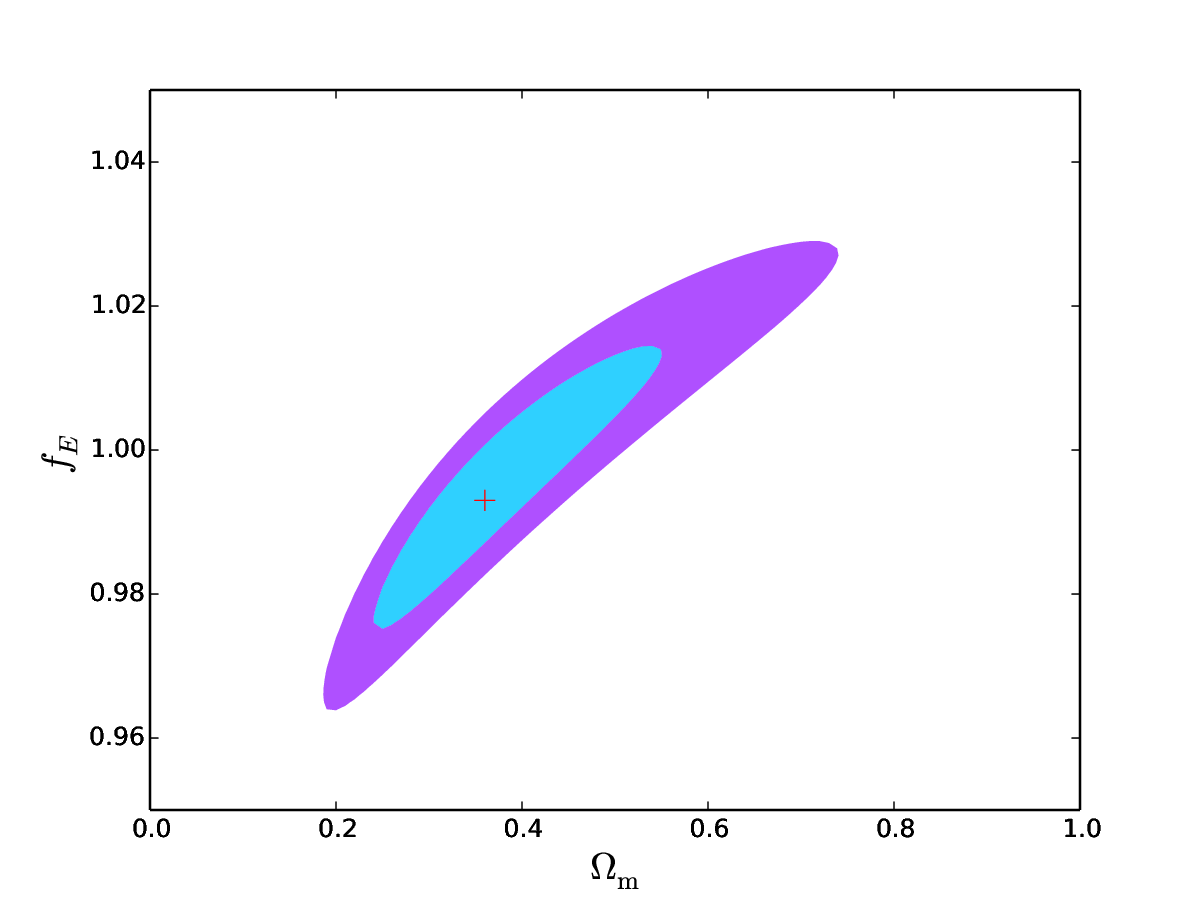}
\includegraphics[width=0.5\textwidth]{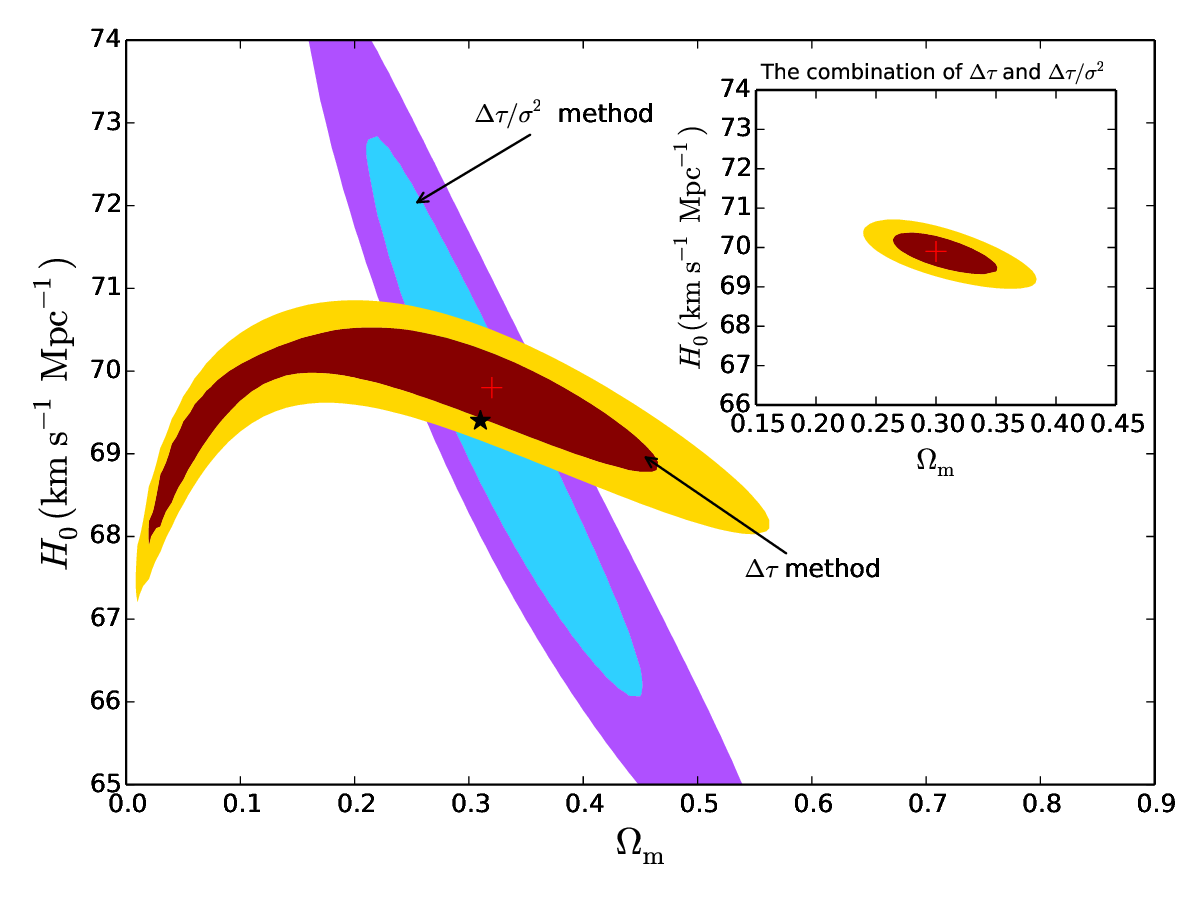}
\caption{The $68.3\%$ and $95.4\%$ confidence regions for
$\Lambda$CDM model in $\Omega_{\rm m}-f_E$ plane (left) for $\sigma$
method and $\Omega_{\rm m}-H_0$ plane (right) for both $\Delta\tau$
and $\Delta\tau/\sigma^2$ methods in the $\Lambda$CDM background.
The samples contain 300, 200 and 150 data points for the methods of
$\sigma$, $\Delta\tau$ and $\Delta\tau/\sigma^2$. The inset of the
right panel shows the constraints from the combination of
$\Delta\tau$ (200 data points) and $\Delta\tau/\sigma^2$(150 data
points).}
\end{figure}

%%%%%%%%%%%%%%%%%%%%%%%%%%%%%%%%% figure 7 and 8 %%%%%%%%%%%%%%%%%%%%%%5
\begin{figure}
\includegraphics[width=0.5\textwidth]{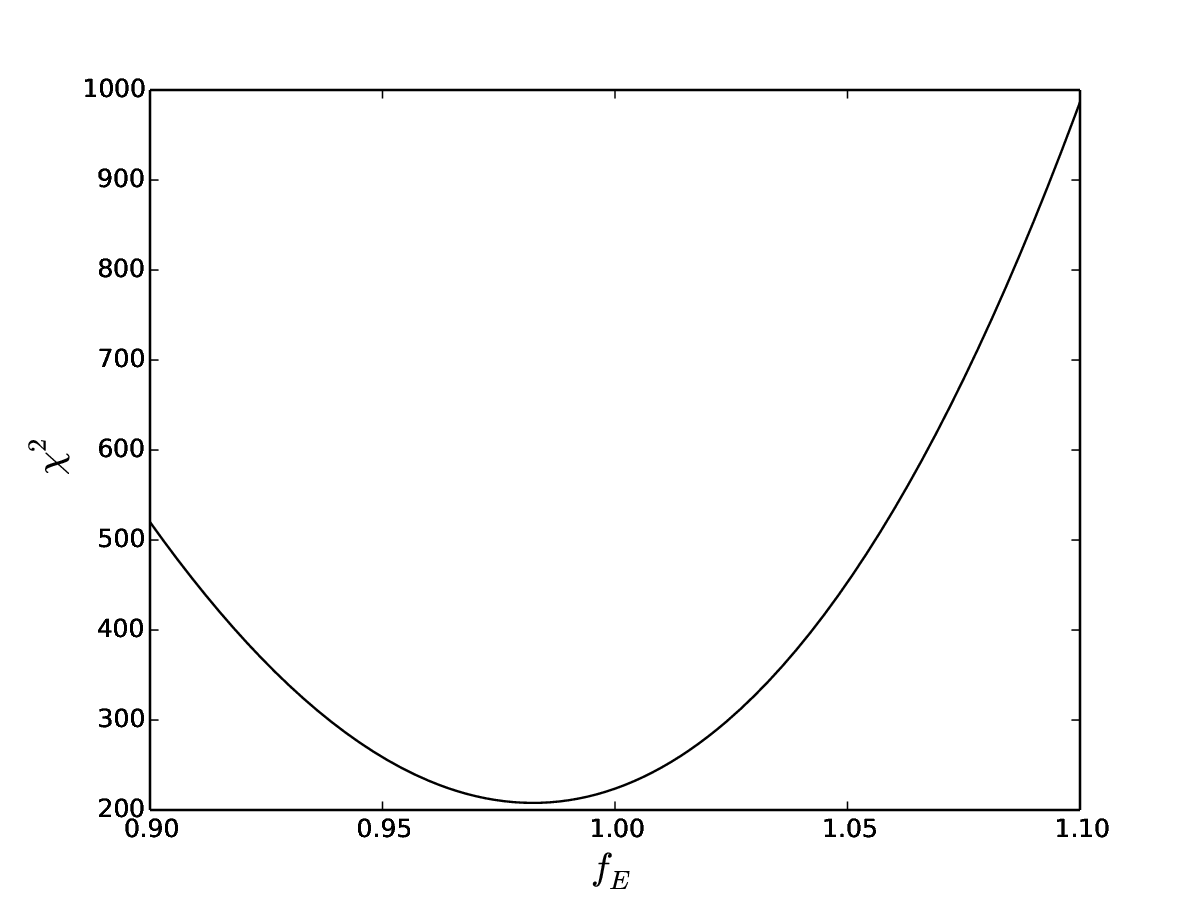}
\includegraphics[width=0.5\textwidth]{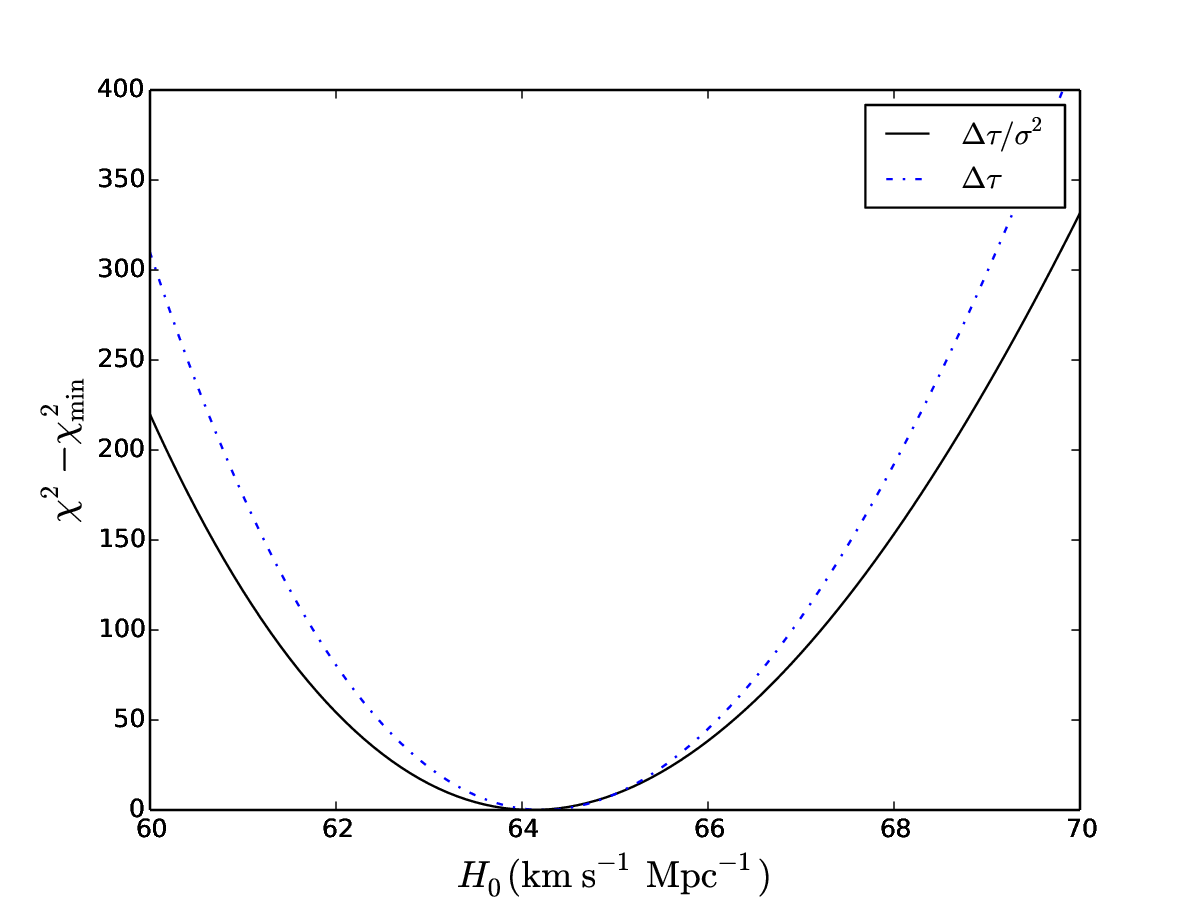}
\caption{$\chi^2$ distributions for $f_E$ and $H_0$ in $R_h=ct$
universe using three different methods: $\sigma$(left),$\Delta\tau$
and $\Delta\tau/\sigma^2$(left). The samples contain 300, 200 and
150 data points for the methods of $\sigma$, $\Delta\tau$ and
$\Delta\tau/\sigma^2$.}
\end{figure}

\begin{figure}
\centering
\includegraphics[width=0.6\textwidth]{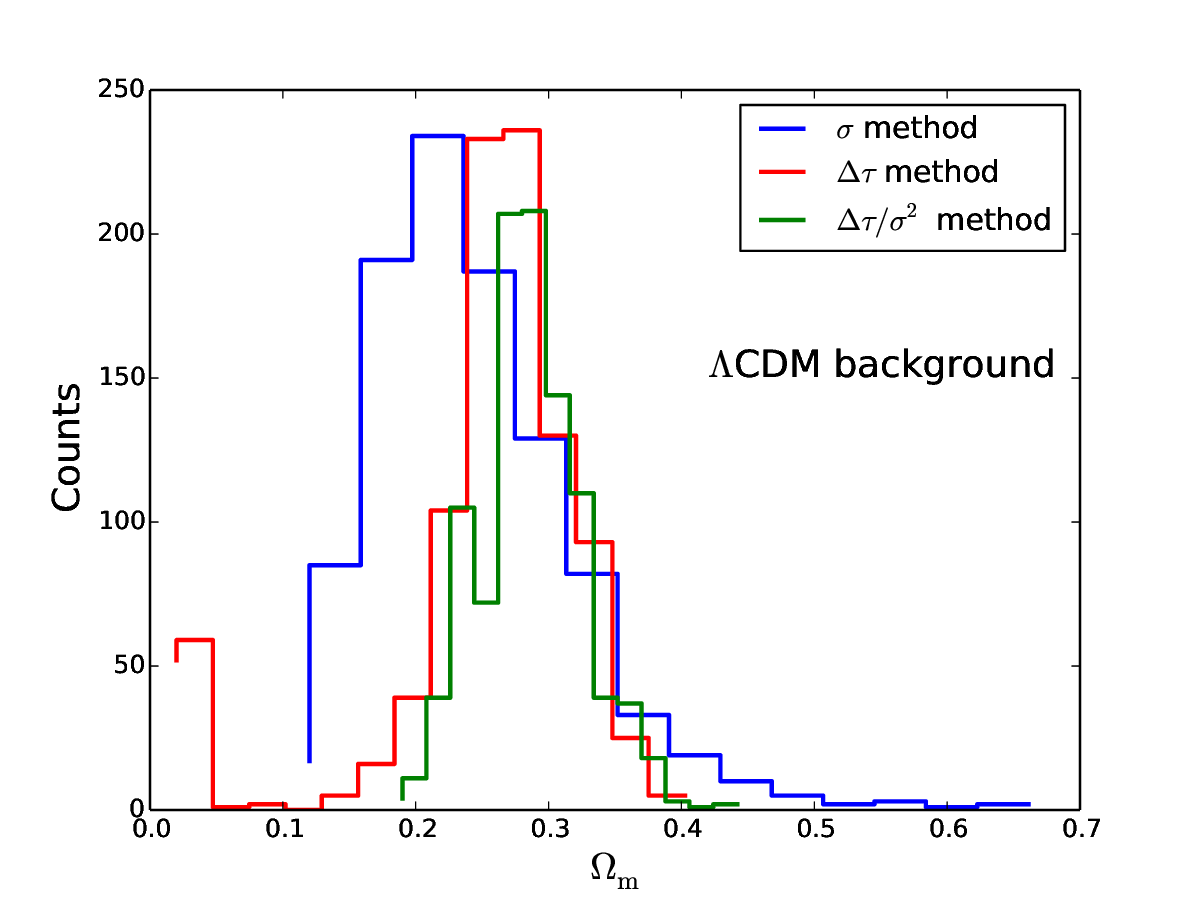}
\caption{Distributions of optimal $\Omega_{m}$ in 1000 repetitive
simulations (N=200) assuming the $\Lambda$CDM background universe.}
\end{figure}

%%%%%%%%%%%%%%%%%%%%%%%%%%%%%%%%%%%5 subsection %%%%%%%%%%%%%%%%%%%%%555
\subsection{$R_h=ct$ Background Cosmology}
We find that a sample of 200 data points for the $\sigma$ method or 600
data points for the $\Delta\tau$ method or $100$ for the combined method
$\Delta\tau/\sigma^2$ is needed independently to rule out the
$\Lambda$CDM model in the $R_h=ct$ background. The results of constraints are
listed in Table 5.

%%%%%%%%%%%%%%%%%%%%%%%%% Table 5 %%%%%%%%%%%%%%%%%%%%%%%%%%%%%%%

\begin{table}
\begin{center}

\begin{tabular}{c|c|c|c}
\hline
Method &Model & Best-fitting parameters$(1\sigma)$& ${\rm BIC}$\\
\hline
\multirow{2}{2.2cm}{$\sigma$($N=200$)}&{$\Lambda$CDM}&$\Omega_{\rm m}=0.50^{+0.43}_{-0.21}, f_E=1.01^{+0.036}_{-0.018}$&249\\
{} &$R_h=ct$&$f_E=0.98^{+0.019}_{0.023}$&238\\
\hline
\multirow{2}{2.2cm}{$\Delta\tau(N=600)$}&{$\Lambda$CDM}&$\Omega_{\rm m}=0.51_{-0.11}^{+0.12},H_0=69.6_{-1.1}^{+0.8} {\rm km\ s^{-1}\ Mpc^{-1}}$ & 670\\
{} &$R_h=ct$&$H_0=71.6_{-0.7}^{+0.9} {\rm km\ s^{-1}\ Mpc^{-1}}$&664\\
\hline
\multirow{2}{2.8cm}{$\Delta\tau/\sigma^2(N=100)$}&{$\Lambda$CDM}&$\Omega_{\rm m}=0.49^{+0.19}_{-0.16}, H_0=71.4^{+3.8}_{-3.2} {\rm km\ s^{-1}\ Mpc^{-1}}$&134\\
{} &$R_h=ct$&$H_0=70.2_{-1.3}^{+1.1} {\rm km\ s^{-1}\ Mpc^{-1}}$&125\\
\hline
\end{tabular}
\caption{Results of one-on-one model comparisons in the $R_h=ct$
background.}
\end{center}
\end{table}
In Figure 9, we illustrate the $68.3\%$ and $95.4\%$ confidence
regions for $\Lambda$CDM model in $\Omega_{\rm m}-f_E$ plane for
the $\sigma$ method and $\Omega_{\rm m}-H_0$ plane for both $\Delta\tau$
and $\Delta\tau/\sigma^2$ methods. In figure 10, we show the
$\chi^2$ distributions of $f_E$ and $H_0$ in $R_h=ct$
model.

Similar conclusions can be obtained in the $R_h=ct$ background. We
also repeat our simulations for 1000 times to differentiate the
constraints on $\Omega_{\rm m}$ utilising three different methods,
the distributions are illustrated in Figure 11 (normal
fitting FWHMs are 0.143, 0.402 and 0.122 for the methods of
$\sigma$, $\Delta\tau$, $\Delta\tau/\sigma^2$ respectively). Noting
that a larger sample (of 600 data points) is needed using the
$\Delta\tau$ method and the FWHM is obviously larger than other
methods. We can draw the conclusion that $\Omega_{\rm m}$ is poorly
constrained using $\Delta\tau$, which is consistent with the
constraint of $\Omega_{\rm m}$ in Figure 4 (left). In addition,
another difference is that $\Lambda$CDM contains more degrees of
freedom to fit the data.

%%%%%%%%%%%%%%%%%%%%%%%%% figure  9 and 10 11 %%%%%%%%%%%%%%
\begin{figure}
\includegraphics[width=0.5\textwidth]{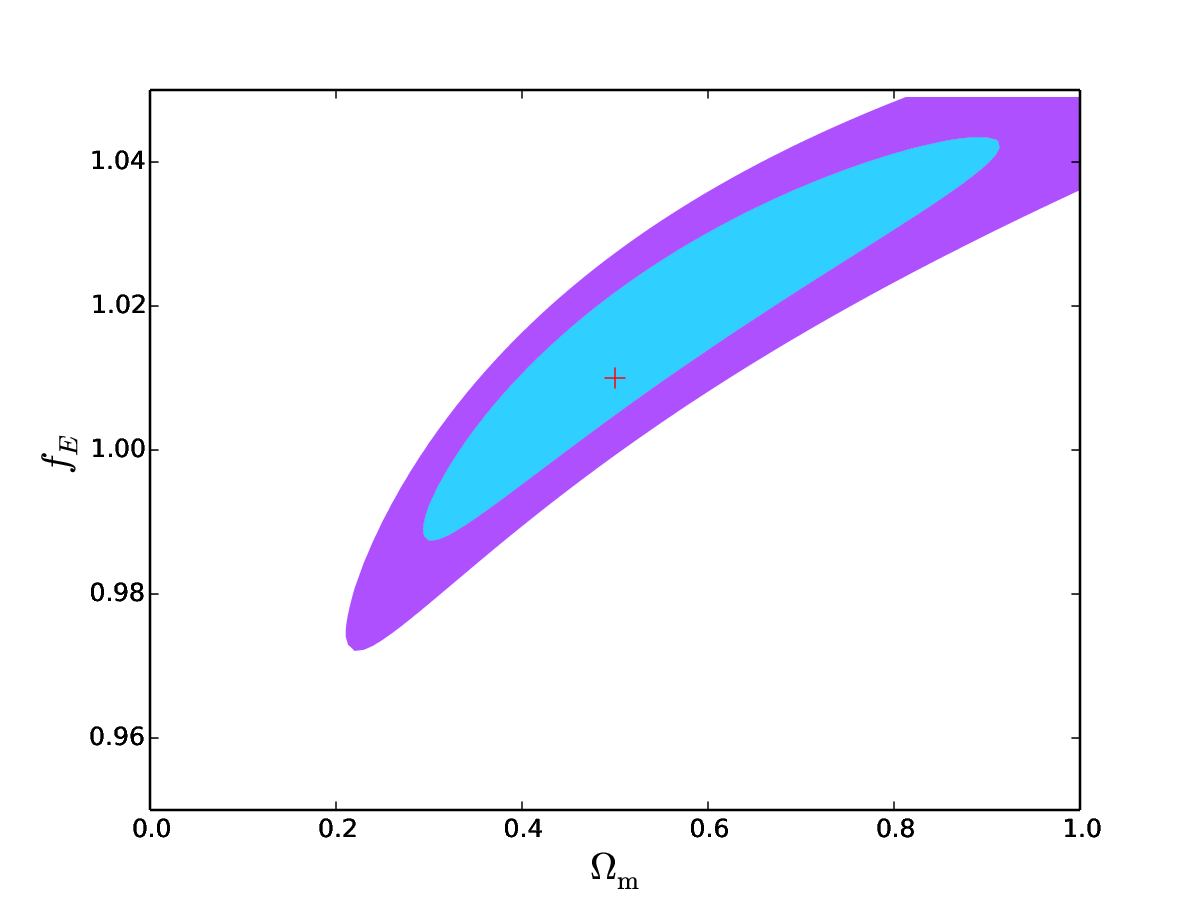}
\includegraphics[width=0.5\textwidth]{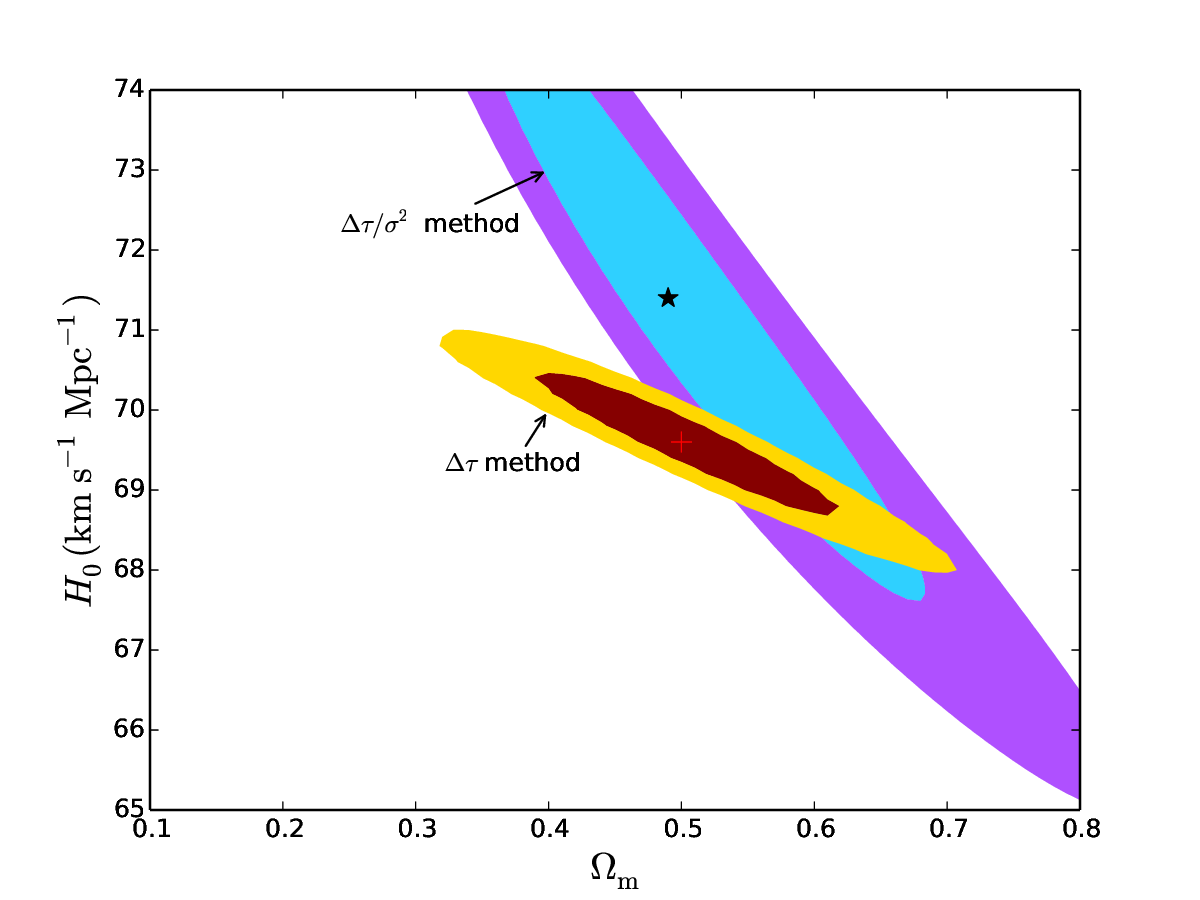}
\caption{The $68.3\%$ and $95.4\%$ confidence regions for the
$\Lambda$CDM model in $\Omega_{\rm m}-f_E$ plane (left) for the $\sigma$
method and $\Omega_{\rm m}-H_0$ plane (right) for both $\Delta\tau$
and $\Delta\tau/\sigma^2$ methods in the $R_h=ct$ background. The
samples contain 300, 200 and 150 data points for the methods of
$\sigma$, $\Delta\tau$ and $\Delta\tau/\sigma^2$, respectively.}
\end{figure}

\begin{figure}
\includegraphics[width=0.5\textwidth]{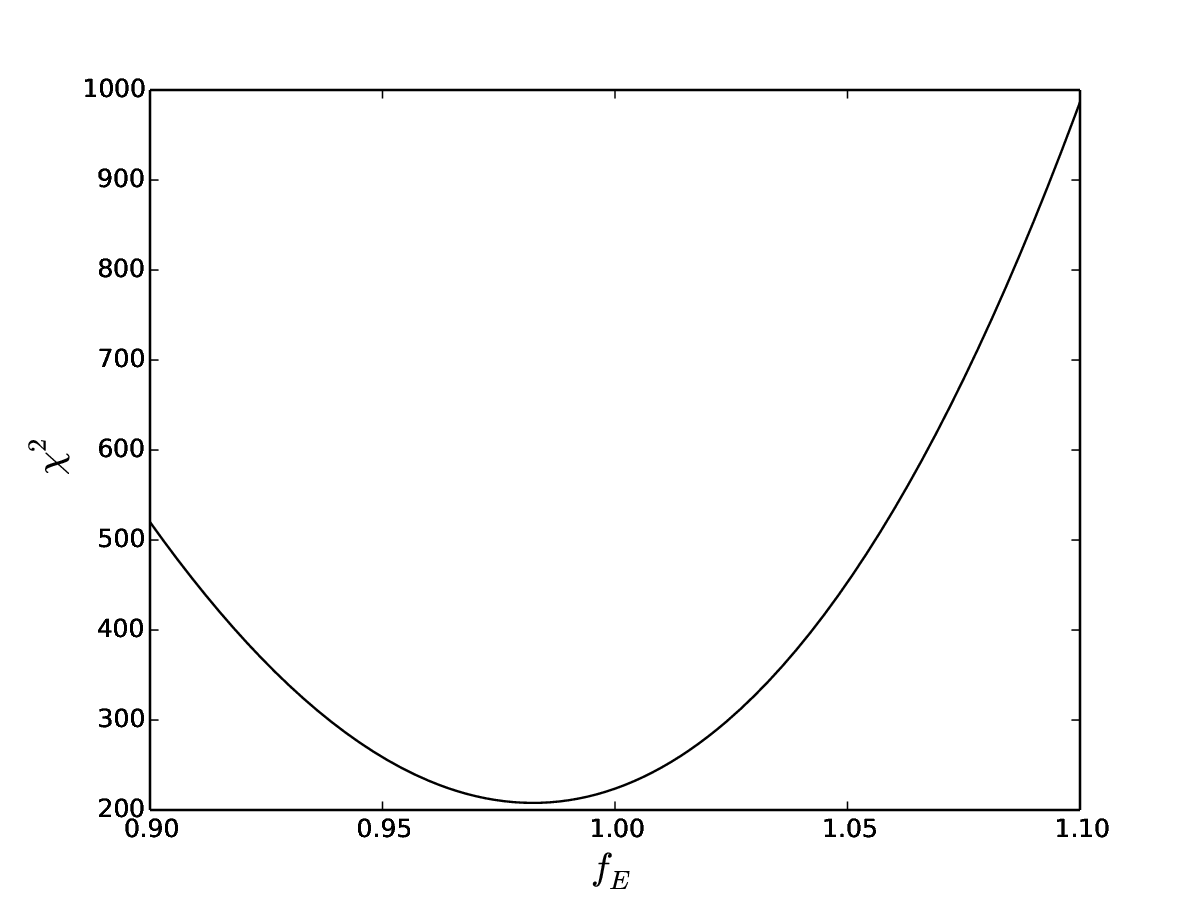}
\includegraphics[width=0.5\textwidth]{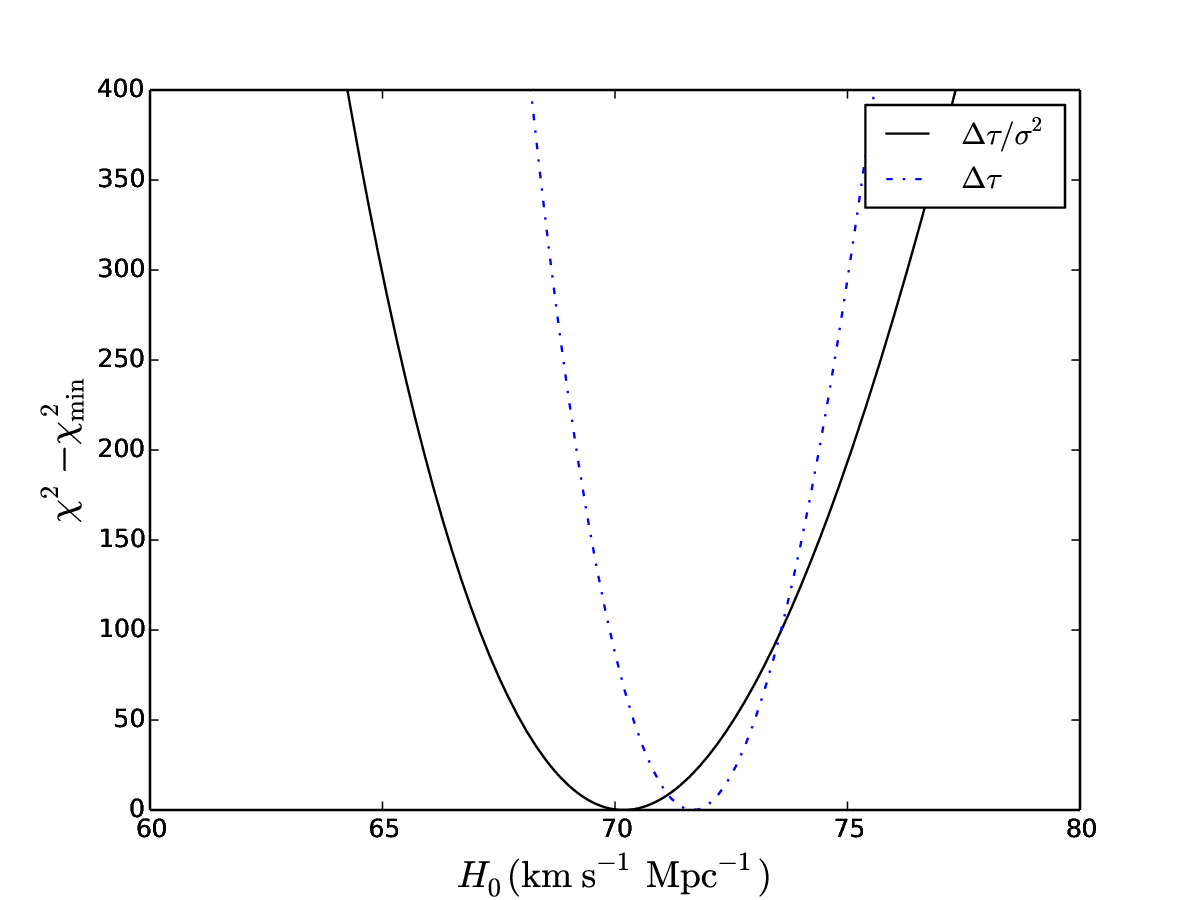}
\caption{$\chi^2$ distributions for $f_E$ and $H_0$ in the $R_h=ct$
universe using three different methods: $\sigma$ (left),$\Delta\tau$
and $\Delta\tau/\sigma^2$ (left). The samples contain 300, 200 and
150 data points for the methods of $\sigma$, $\Delta\tau$ and
$\Delta\tau/\sigma^2$, respectively.}
\end{figure}

\begin{figure}
\centering
\includegraphics[width=0.6\textwidth]{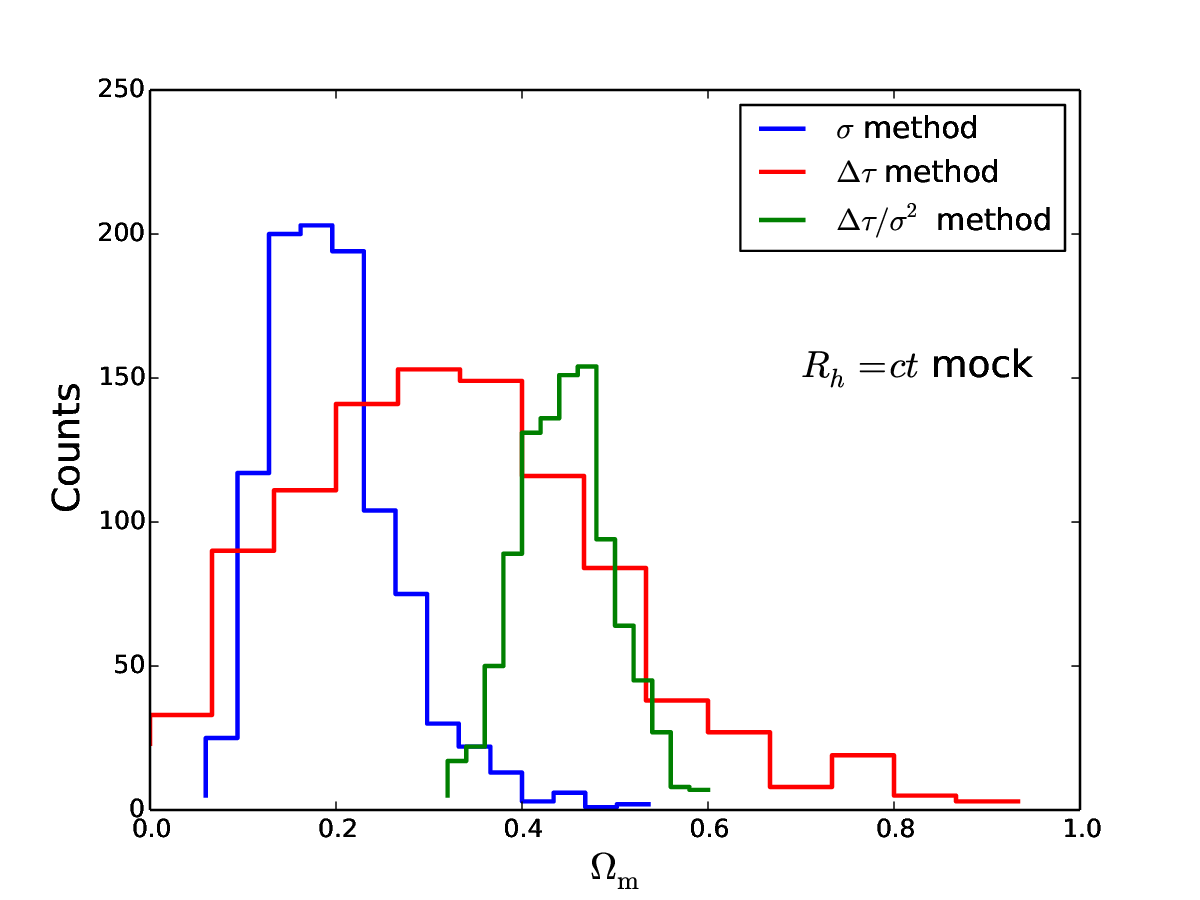}
\caption{Distributions of optimal $\Omega_{\rm m}$ in 1000
repetitive simulations (N=200) assuming the background is the
$R_h=ct$ model.}
\end{figure}
%%%%%%%%%%%%%%%%%%%%%%%%%%%%%% subsection 4.3 %%%%%%%%%%%%%%%%%%%%5
\subsection{Independence Tests for $\sigma$, $\Delta\tau$ and $\Delta\tau/\sigma^2$}
The previous discussions in this section are based on the assumption
that these three methods are independent. This assumption is
appropriate to differentiate the efficiency of different methods.
However, when we put the $\Delta\tau/\sigma^2$ method into practical
cosmological tests, we must consider its independence on the other
two methods. Here, independence tests are performed using MC
simulations and the steps are given as follows.

(i) Generate two samples (200 data points in each sample)
for $\sigma$ and the $\Delta\tau$ methods using the previous scheme
in the beginning of this section assuming the background is
$\Lambda$CDM. The lensing redshifts ($z_l$) of the corresponding
data points in each sample should be the same, so as $z_s$.

(ii) In this step we generate the sample for the
$\Delta\tau/\sigma^2$ method. Here, $z_l$, time-delay $\Delta\tau$
and velocity dispersion $\sigma$ can be obtained directly from the
samples in step (i). Noting that $\Theta=\theta_B^2-\theta_A^2$ and
$\theta_E=(\theta_A+\theta_B)/2$, $\Delta\theta=\theta_B-\theta_A$
can be expressed in the form $\Theta/(2\Delta\theta).$ That means we
can combine the three samples as one with the measurement of both
time-delay, velocity dispersion, $\theta_A$ and $\theta_B$.

(iii) Using the samples generated in (i) and (ii) to
constrain $\Omega_{\rm m}$ and obtain the optimal $\Omega_{{\rm m}}$
for these three methods.

(iv) Repeat steps (i)-(iii) $n$ times (here, $n$=500) and
get three samples of optimal $\Omega_{\rm m}$ for the methods of
$\Delta\tau$, $\sigma$ and $\Delta\tau/\sigma^2$ respectively.

Figure 12 shows the correlations of the obtained $\Omega_{\rm m}$
samples. The $x$-axis is the sample of $\Omega_{{\rm m},1}$ obtained
from one method while $y$-axis is the sample of $\Omega_{{\rm m},2}$
obtained from another method, for instance, $\Omega_{{\rm m},1}$
(the $\sigma$ method) versus $\Omega_{{\rm m},2}$ (the
$\Delta\tau/\sigma^2$ method). From Figure 12 we find the samples of
optimal $\Omega_{\rm m}$ obtained from $\Delta\tau/\sigma^2$ and
$\sigma$ are strongly and positively correlated, which means that
resultant $\Omega_{\rm m}$ from the $\Delta\tau/\sigma^2$ method and
the method of $\sigma$ are not independent. Besides, this figure
illustrates that there is no obvious correlation between
$\Delta\tau$ and $\sigma$, $\Delta\tau/\sigma^2$ and $\Delta\tau$.
Despite the independence tests of these methods revealing a
correlation between $\Delta\tau/\sigma^2$ and $\sigma$,
$\Delta\tau/\sigma^2$ could be considered as an improved method of
$\sigma$, especially for the lensing systems with the measurement of
both time-delays, velocity dispersions and the radii of two images
($\theta_A$ and $\theta_B$).
%%%%%%%%%%%%%%%%%%% Figure 12 %%%%%%%%%%%%%%%%%%%%%%
\begin{figure}
\centering
\includegraphics[width=0.6\textwidth]{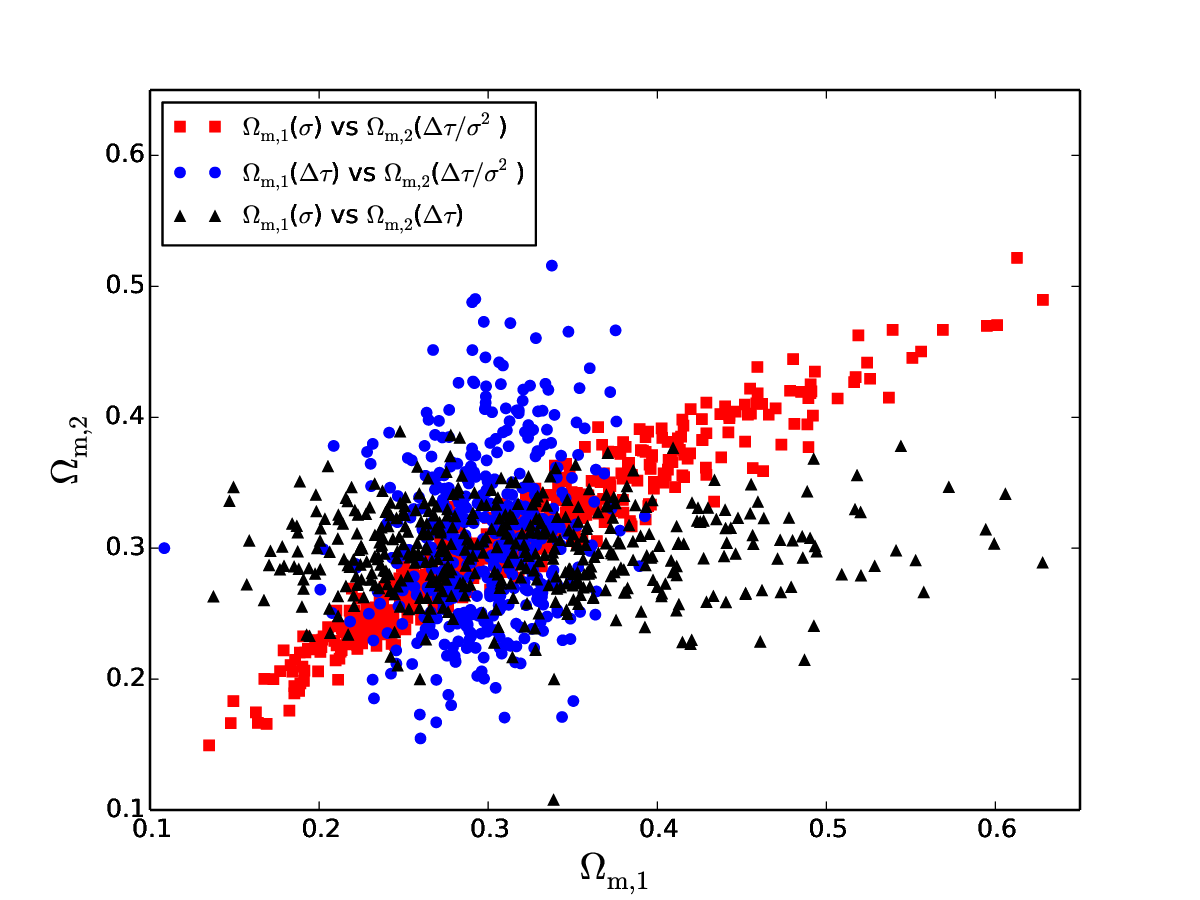}
\caption{Correlations of obtained $\Omega_{\rm m}$ samples using
different methods: $\sigma$ versus $\Delta\tau/\sigma^2$(red
square), $\Delta\tau$ versus $\Delta\tau/\sigma^{2}$(blue circle)
and $\sigma$ versus $\Delta\tau$ (black triangle)}
\end{figure}

\section{Conclusions and discussions}
In this paper, we use three methods to constrain cosmological
parameters and make one-on-one comparisons between $\Lambda$CDM
Universe and $R_h=ct$ Universe. Using sample I containing 36
two-image Einstein circle lenses, we find the current two-image data
set failed to rule out one Universe at the $99.7\%$ confidence
level. In addition, we use a sample of 12 time-delay lensing systems
to compare $\Lambda$CDM and $R_h=ct$. By using Akaike Information
Criterion we find $R_h=ct$ is superior to $\Lambda$CDM with a
likelihood of $74.8\%$. More data points are required to rule out
one model at a higher confidence level.

Since lack of gravitational lensing systems observed with both
$\sigma$, $\Delta\tau$ and $\Delta\theta=\theta_B-\theta_A$, the
sample for $\Delta\tau/\sigma^2$ can merely be obtained through
simulations. Therefore, we use MC simulations to compare different
methods concerning velocity dispersion $\sigma$, time-delay
$\Delta\tau$ and their combination, $\Delta\tau/\sigma^2$, in
one-on-one comparisons. Through assuming a background universe, we
try to find the least number of data points to rule out another
cosmological model at a $99.7\%$ confidence level. From the
distributions of optimal $\Omega_{\rm{m}}$ we find that the
$\Delta\tau/\sigma^2$ is superior to $\Delta\tau$ in the constraints
of $\Omega_{\rm m}$.

In order to differentiate the efficiency of different methods, we
repeat our simulations for 1000 times to compare the constraints on
$\Omega_{\rm m}$ \textbf{utilising} three different methods. In the
simulation, we assign the number of data points in each sample to be
200. For both backgrounds, we find that $\Delta\tau/\sigma^2$ can
give a tighter constraint on $\Omega_{\rm m}$ than $\sigma$ and
$\Delta\tau$. As shown in Figure 13, we plot the best-fit data
points in $\Omega_{\rm m}-H_0$ plane for $\Delta\tau$ and
$\Delta\tau/\sigma^2$ methods. The only difference of
$\Delta\tau/\sigma^2$ in $\Lambda$CDM Universe and $R_h=ct$ Universe
backgrounds is the shift of the best-fitting $\Omega_{\rm m}$ and
$H_0$. However the distribution of optimal $\Omega_{\rm m}$ obtained
through $\Delta\tau$ in the $R_h=ct$ Universe background is more
diffuse compared with $\Lambda$CDM background. This can explain that
the sample needed in the method of $\Delta\tau$ in the $R_h=ct$
background is much larger than in the $\Lambda$CDM background (600
data points versus 200 data points).

These three methods are useful to compare cosmological models and
each of them has its advantages in special aspects. Although
$\Delta\tau/\sigma^2$ and $\sigma$ are not independent, it can be
considered as an improved method of $\sigma$. Besides, from our
independence tests we find that the $\Delta\tau$ method and
$\Delta\tau/\sigma^2$ are independent, thus the joint consideration
of them can be used to give a tight constraint in $\Omega_{\rm
m}-H_0$ plane for $\Lambda$CDM model. Despite the relative lack of
observational data, future studies of lensing systems and high
resolution observations of galaxies will provide more geometry and
dynamic information about strong gravitational lenses. Then, the
$\Delta\tau/\sigma^2$ method will become a powerful method in
cosmological model selections.

%%%%%%%%%%%%%%%%%%%%%%%%%%%%%%%%%% figure 13 %%%%%%%%%%%%%%%%%%%%%555
\begin{figure}
\centering
\includegraphics[width=0.6\textwidth]{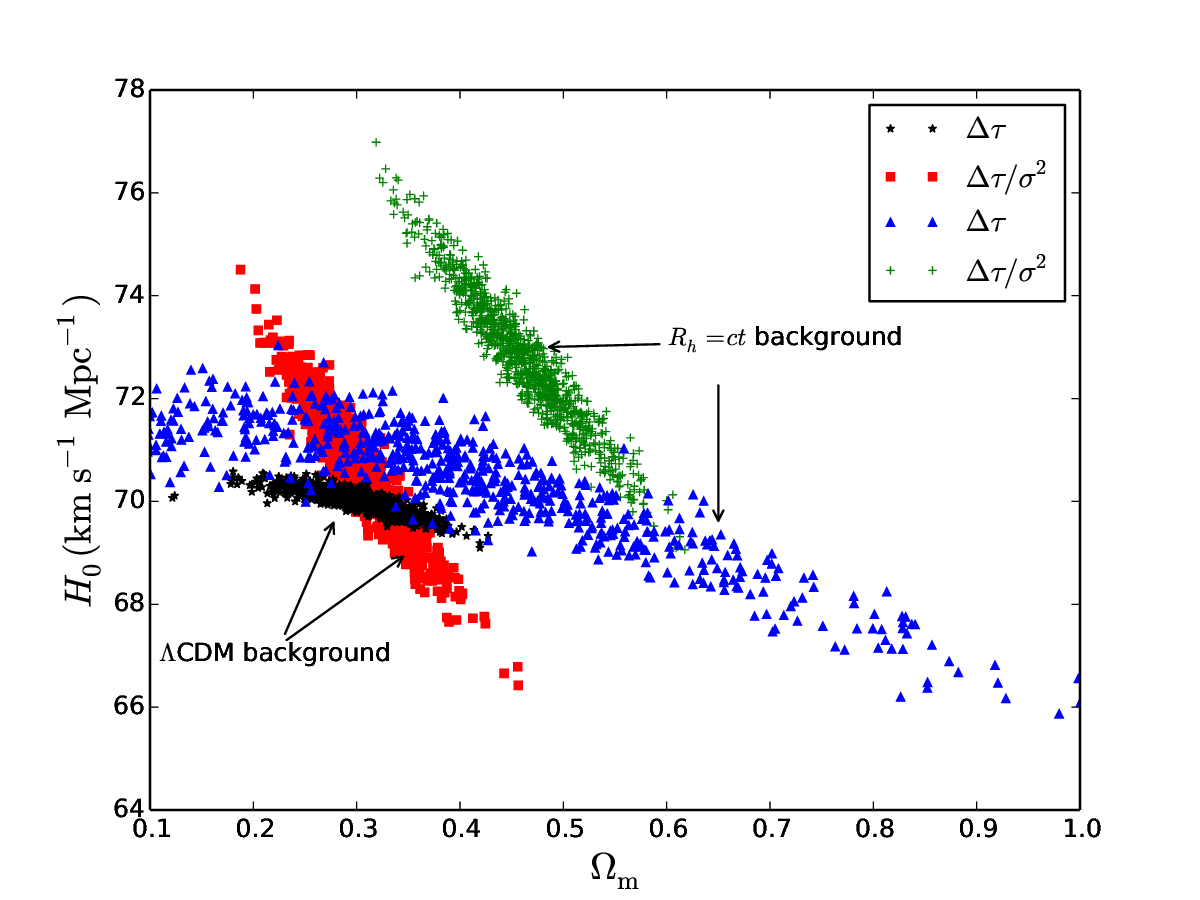}
\caption{Distributions of best-fitting points using the method of
$\Delta\tau$ or $\Delta\tau/\sigma^2$ in $\Omega_{\rm m}-H_0$ plane
using 1000 repetitive simulations (N=200) assuming different
background models.}
\end{figure}

\section*{Acknowledgements}
We thank Shun-Sheng Li for fruitful discussions and careful reading
of the manuscript. This work is supported by the National Basic
Research Program of China (973 Program, grant No. 2014CB845800) and
the National Natural Science Foundation of China (grants 11422325,
11373022, J1210039 and 11033002), the Excellent Youth Foundation of
Jiangsu Province (BK20140016). C.C.Y is also supported by Innovation
Program of Undergraduates, Ministry of Education of China under
Grant No. S1410284047. 

\end{document}